\documentclass[fleqn,usenatbib]{mnras}

\usepackage{newtxtext,newtxmath}
\usepackage{float}

\usepackage[T1]{fontenc}

\DeclareRobustCommand{\VAN}[3]{#2}
\let\VANthebibliography\thebibliography
\def\thebibliography{\DeclareRobustCommand{\VAN}[3]{##3}\VANthebibliography}

\usepackage{graphicx}	
\usepackage{amsmath}	

\title[Accretion properties of UGC~6728]{The accretion properties of a low-mass Active Galactic Nucleus: UGC~6728}

\author[Nandi et al.]{
Prantik Nandi,$^{1}$\thanks{E-mail: prantiknandi007@gmail.com}
Sachindra Naik,$^{1}$
Arka Chatterjee,$^{2}$
Sandip K. Chakrabarti,$^{3}$
Samar Safi-Harb$^{2}$
\newauthor
and  Neeraj Kumari$^{1}$
\\
$^{1}$Astronomy and Astrophysics Division, Physical Research Laboratory, Ahmedabad-380009, India\\
$^{2}$Department of Physics \& Astronomy, University of Manitoba, Winnipeg, R3T 2N2, Canada \\
$^{3}$Indian Centre for Space Physics, Netaji Nagar,  Kolkata, 700099, India \\
}

\date{Accepted XXX. Received YYY; in original form ZZZ}

\pubyear{2023}

\begin{document}
\label{firstpage}
\pagerange{\pageref{firstpage}--\pageref{lastpage}}
\maketitle

\begin{abstract}
We present a comprehensive analysis of approximately $15$ years ($2006-2021$) of X-ray observations of UGC~6728, a low-mass bare AGN, for the first time. Our study encompasses both spectral and temporal aspects of this source. The spectral properties of this source are studied using various phenomenological and physical models. We conclude that (a) the observed variability in X-ray luminosity is not attributed to the Hydrogen column density ($N_H$) as UGC~6728 exhibits a bare nucleus, implying a negligible $N_H$ contribution along the line of sight, and (b) the spectral slope in the X-ray band demonstrates a systematic variation over time, indicating a transition from a relatively hard state to a comparatively soft state. We propose that the underlying accretion dynamics around the central object account for this behavior. By performing X-ray spectral fitting, we estimate the mass of the central supermassive black hole (SMBH) in UGC~6728 to be $M_{BH}=(7.13\pm1.23)\times10^5$ M$_\odot$  with spin $a=0.97^{+0.20}_{-0.27}$ and inclination angle $i=49.5\pm14.5$ degree. Based on our spectral and temporal analysis, we suggest that UGC~6728 lacks a prominent Compton hump or exhibits a very subtle hump that remains undetectable in our analysis. Furthermore, the high-energy X-ray photons in this source are likely to originate from the low-energy X-ray photons through inverse Compton scattering in a Compton cloud, highlighting a connection between the emission in two energy ranges. We notice a strong soft excess component in the initial part of our observations, which later reduced substantially. This variation of soft excess is explained in view of accretion dynamics.

\end{abstract}

\begin{keywords}
galaxies: active – galaxies: Seyfert – X-rays: galaxies – X-rays: individual: UGC~6728.
\end{keywords}



\section{Introduction}
Active galactic nuclei (AGNs) are among the most luminous sources which are powered by the process of accretion of matter from the host galaxy \citep{Storchi2019} onto the supermassive black holes (SMBHs) \citep{Lynden-Bell1969, Rees1984} located at their center. The mass of an SMBH ranges from $10^5$ to $10^9$ M$_\odot$ \citep{Kormendy1995, Peterson2004, EHT2019}. The AGNs are generally characterized by their complex spatial and temporal variabilities  \citep{McHardy1999, Nandra2001, Webb2000, Ballantyne2014, Zoghbi2017} over the entire range of the electromagnetic spectrum, starting from radio to very high energy $\gamma-$rays. The X-ray photons are believed to be originated from a region closer to the central SMBH \citep{Fabian1989, Wilkins2021}. Thus, X-ray spectroscopy is a powerful tool to probe the innermost regions of the accretion disc around the SMBHs. It allows us to determine the physical conditions of the plasma surrounding the central X-ray source. By observing these X-ray photons, we can characterize the fundamental parameters of the black hole, such as its mass, spin, etc. These parameters play a crucial role in understanding the physical nature of the central engine and its cosmological evolution \citep{Fabian2012, Kormendy2013}. However, our knowledge of the physical processes in AGNs is mostly based on the galaxies which harbour a SMBH of mass greater than $10^6$ M$_\odot$. However, the low mass AGNs $(M_{BH} < 10^6 M_\odot)$ are yet to be explored. One can explore the missing link to probe whether the accretion mechanism is independent of the mass of the black hole. 

According to standard accretion disc theory \citep{Pringle1973, Shakura1973}, X-ray photons are thought to be originated from the innermost region of the accretion disc, which surrounds the central black hole. Observationally, the X-ray spectra of AGNs are explained by the power-law continuum \citep{Bianchi2009, Sobolewska2009} with high-energy cut-off \citep{Fabian2015, Fabian2017, Tortosa2018, Middei2019, Balokovic2020}. The power-law continuum is thought to be generated by inverse Comptonization \citep{Sunyaev1980} of thermal optical/UV photons in a hot corona or Compton cloud located close to the accretion disc \citep{Haardt1991}. However, the geometry and location of the Compton reprocessing region are still not well understood. This cloud can be located above the disc \citep{Haardt1991, Haardt1993, Poutanen1996} or at the base of the jet \citep{CT1995, Fender1999, Markoff2005}. This region is formed by the hot ($T\sim 10^9 $ K) and radiatively inefficient plasma \citep{Sunyaev1980}, which behaves like a quasi-Bondi flow \citep{Ichimaru1977}. 

In addition to the power-law continuum, the other spectral features, such as absorption \citep{Turner1997, Risaliti1999, Turner2009}, reflection \citep{Ross1993, Ross2005, Garcia2013, Garcia2014}, soft X-ray excess emission \citep{Halpern1984, Singh1985, Arnaud1985, Miniutti2004, Crummy2006, Walton2013, Liebmann2018, Nandi2021, Nandi2023}, Fe lines \citep{Ross2005, Fabian2009, Garcia2010} and a high-energy cut-off \citep{Sunyaev1980, Haardt1991, Haardt1993} are also present in the X-ray spectra of the AGNs. The absorption effect on the X-ray spectrum is produced by the equivalent Hydrogen column density along the line of sight \citep{Antonucci1985, Netzer2013, Jana2020}. We often observe an excess emission in the soft X-ray (below $2$ keV) energy band, known as `soft-excess' \citep{Halpern1984, Singh1985, Arnaud1985, Magdziarz1998, Jin2012, Done2012, Petrucci2013, Petrucci2018, Kubota2018, Middei2020, Nandi2021}. This feature is mostly observed in Seyfert~1 AGNs \citep{Pravdo1981, Arnaud1985, Singh1985, Turner1989}, a subclass of radio-quiet AGNs. The X-ray spectra below $10$ keV are often associated with several fluorescent emission lines. Among them, the most prominent is the Fe K$_\alpha$ line at $\sim6.4$ keV \citep{Pounds1990, Ross2005, Fabian2009, Garcia2010}. From the albedo calculation of the accretion disc, the Comptonized photons produce the reflection component \citep{CT1995}. The signature of this component is visible in $15$ to $50$ keV range with a peak at around $\sim30$ keV, which is called the `Compton hump' \citep{Krolik1999}. 

 In this scenario, X-ray observations play a crucial role in constraining the geometry, such as the size and location, of the emitting region with a proper understanding of its formation. A common characteristic observed in X-ray studies of nearby bright Seyfert galaxies is the detection of X-ray reverberation lags. This phenomenon suggests that the X-ray emitting region is compact and lies within a few tens of the gravitational radii of the black hole \citep{Fabian2009, De2013, Cackett2014, Emmanoulopoulos2014, Uttley2014, Kara2016}. The spectral modelling of optical to X-ray spectra from AGNs also yields a similar size of X-ray emitting region \citep{Petrucci2013, Done2013, Kara2013, Nandi2019, Nandi2021}.  From a physical perspective, the two-component advective flow (TCAF) \citep{CT1995} model may provide a self-consistent picture of the accretion dynamics around the central SMBHs. This model is quantified by four flow parameters: two types of accretion flow rates, namely, disc rate $(\Dot{m}_d)$ \citep{Shakura1973} and halo rate $(\Dot{m}_h)$ \citep{Chakrabarti1989, Chakrabarti1990, Chakrabarti1995} in the unit of Eddington rate $(\Dot{M}_{Edd})$, the dimension of the Compton cloud or CENBOL through shock location $(X_s)$ in the unit of Schwarzschild radius $(R_s=2GM_{BH}/c^2)$, and the density of this cloud by the shock compression ratio $(R)$ \citep{Chakrabarti1989}, which is defined as the ratio of the post-shock and the pre-shock flow densities. Along with these parameters, we also need an intrinsic parameter, namely, the mass of the central black hole (in the unit of M$_\odot$) and an extensive parameter, called normalization, which is used to match the observed spectrum with the theoretical one, and essentially a function of the distance of the source and the collecting area of the instrument. This model is used to constrain the geometry around the black hole mass from few solar mass \citep{Chakrabarti1997, Dutta2010, Debnath2014} to supermassive AGNs \citep{Mandal2008, Nandi2019, Nandi2021}. However, for the reflection-dominated AGN spectra, various geometries for the Compton cloud have been proposed for the X-ray spectra fitting \citep{Dauser2013, Garcia2014, Niedzwiecki2016, Ingram2019, Mastroserio2021, Doviak2022}. It is to be noted that the TCAF model includes the reflection component in the very process of obtaining the solution \citep{CT1995}.

UGC~6728 is a relatively less explored nearby ($z$=0.0065\footnote{\url{https://ned.ipac.caltech.edu/byname?objname=UGC+6728&hconst=67.8&omegam=0.308&omegav=0.692&wmap=4&corr_z=1}} \citep{Kalberla2005}) Seyfert~1 AGN, which shows broad emission lines in the optical band \citep{Bentz2016}. This AGN is a late-type galaxy and contains an SMBH of mass M$_{BH}=(7.1\pm4.0)\times10^5$ M$_\odot$ \citep{Bentz2016}. The X-ray observation of UGC~6728 reported that it has a `bare nucleus,' i.e., little or no obscuration along the line of sight to the nucleus \citep{Walton2013}.  

We organize the paper as follows. In Section~\ref{sec:obs}, we describe the details of the data selection and reduction techniques for different observatories. Then in Section~\ref{sec:result}, we perform our analysis of X-ray data from various observations and present the corresponding results. We further divide our analysis into two sub-sections. In Section~\ref{sec:timing}, we analyze and discuss the timing properties of corresponding observations. Section~\ref{sec:spectral} describes the various phenomenological and physical models to extract different physical properties of this source. Later, we discuss our findings from the X-ray data analysis of this source in Section~\ref{sec:discussions}. Finally, we draw our conclusions based on our results in Section~\ref{sec:conclusions}.


\section{Observations and Data Reduction}
\label{sec:obs}
In the present work, we use publicly available archival data of the low mass, bare AGN UGC~6728 from {\it XMM-Newton}, {\it Swift}, and {\it NuSTAR} using HEASARC\footnote{\url{http://heasarc.gsfc.nasa.gov/}}. We reprocessed all the data using {\tt HEAsoft v6.26.1} \citep{Arnaud1996}, which includes {\tt Xspec v12.10.1f}.

\subsection{ NuSTAR}
{\it NuSTAR} is a hard X-ray focusing telescope that operates in the energy range from $3.0$ to $79.0$ keV using two identical moduli, FPMA and FPMB \citep{Harrison2013}. This telescope observed UGC~6728 in two phases in 2016 and 2017. The details of the observations are listed in Table~\ref{tab:1}. We obtained the observational data from NASA's HEASARC archive\footnote{\url{https://heasarc.gsfc.nasa.gov/cgi-bin/W3Browse/w3browse.pl}} and used standard data reduction technique\footnote{\url{https://heasarc.gsfc.nasa.gov/docs/nustar/analysis/}} to reprocess the data. The data were reprocessed by {\it NuSTAR} Data Analysis Software {\tt NUSTARDASv1.4.1}. We used standard filtering criteria to generate clean event files using the {\tt nupipeline} task. We used the latest calibration data (20230918) from the {\it NuSTAR} database\footnote{\url{https://heasarc.gsfc.nasa.gov/FTP/caldb/data/nustar/fpm/}} to calibrate the data. The source and the background products were extracted by considering circular regions of $60$ arcsec and $120$ arcsec radii, respectively. We considered the centre of the source region to be at the source coordinates. The background region was considered far from the source region to avoid any contamination. Later, we used the {\tt nuproduct} task to extract the source spectra and rebinned them by ensuring at least 50 counts per bin by using the {\tt GRPPHA} task.

\subsection{ XMM-Newton}
UGC~6728 was observed with {\it XMM-Newton} \citep{Jansen2001} only once on 23 February 2006. The {\it XMM-Newton} observation of the source was carried out in {\tt Full Window Mode}. The details of the observation are given in Table~\ref{tab:1}. We used the Science Analysis System ({\tt SAS v16.1.0}\footnote{\url{https://www.cosmos.esa.int/web/xmm-newton/sas-thread-epatplot}}) to reprocess the raw data of EPIC-pn \citep{Struder2001} with calibration data files dated 2 February 2018. To prevent the edge of CCD and the bad pixels, we used only the unflagged events using {\tt FLAG == 0}. Along with this, we also used {\tt PATTERN $\leq$ $4$} for single and double pixels. We exclude the photon flares by considering appropriate Good Time Interval ({\tt GTI}) files to acquire the maximum signal-to-noise ratio. For the spectrum, initially, we used circular regions of $30$ arcsec and $60$ arcsec radii for the source and background, respectively. Then we apply the SAS task {\tt especget} for spectrum extraction. After that, we inspected the spectrum for photon pile-up using SAS task {\tt epaplot}. To remove the pile-up effect, we considered numerous values of the inner radii for the source and investigated the presence of pile-up\footnote{\url{https://www.cosmos.esa.int/web/xmm-newton/sas-thread-epatplot}}. Finally, we found that in an annular region with inner and outer radii of $10$ arcsec and $30$ arcsec, respectively, the source spectrum became free from the pile-up effect. We use this annular region to extract the final source spectrum and use it to fit various spectral models. The response files ({\it arf} and {\it rmf} files) were generated with the SAS task {\tt arfgen} and {\tt rmfgen}, respectively. Finally, we rebinned the data by ensuring at least $100$ counts per bin using {\tt GRPPHA} on the final $0.2$ to $10.0$ keV EPIC-pn spectrum.

\subsection{ Swift}
{\it Swift}/XRT \citep{Burrows2005} is an X-ray focusing telescope sensitive in the energy range of $0.3$ to $10.0$ keV. UGC~6728 was monitored with this telescope in three epochs from 2006 to 2021. Between them, this source was observed simultaneously with {\it NuSTAR} in 2016 and 2017 (see Table~\ref{tab:1} for details). To generate X-ray spectra between $0.3$ to $10.0$ keV, we used standard online tools provided by the UK {\it Swift} Science Data Centre\footnote{\url{https://www.swift.ac.uk/user_objects/}} \citep{Evans2009}. In the case of 2021 observations, we combined all the spectra into a single spectrum to achieve enough exposure time. We also considered these observations individually in our spectral fitting using a simple power-law model. However, we did not observe any significant difference in the model parameters. Finally, we used {\tt GRPPHA} task to rebin the XRT spectra, ensuring a minimum of 10 counts per bin.

\begin{table}
\caption{Observation Log.}
\label{tab:1}
\begin{tabular}{ccccc}
\hline
ID&Date           & Obs. ID     & Instrument              & Exposures   \\
  &(yyyy-mm-dd)   &             &                          & (ks)      \\
\hline
&&&&\\
XMM & 2006-02-23 & 0312191601 & {\it XMM-Newron}/epic-pn & 11.8               \\
&&&&\\
SU  & 2009-06-06 & 704029010  & {\it Suzaku}/XIS-HXD     & 46~\&~39$^\dagger$ \\
&&&&\\
XRT1  &  2016-07-10 & 00081098001 & {\it Swift}/XRT          & 06.9  \\
&&&&\\
N1    &  2016-07-10 & 60160450002 & {\it NuSTAR}/FPMA-FPMB   & 22.6  \\
&&&&\\
XRT2  &  2017-10-13 & 00088256001 & {\it Swift}/XRT          & 06.9  \\
&&&&\\
N2    &  2017-10-13 & 60376007002 & {\it NuSTAR}/FPMA-FPMB   & 58.1  \\
&&&&\\
XRT3  &  2020-09-05 & 00013662001 & {\it Swift}/XRT          & 01.8  \\
&&&&\\
XRT4  &  2021-04-13 & 00013662002 & {\it Swift}/XRT          & 02.2  \\
      & -2021-10-25 & 00096132008 &           \\
&&&&\\
\hline
\end{tabular}
$^\dagger$ exposure time for XIS and HXD, respectively.
\end{table}

\subsection{ Suzaku}
 {\it Suzaku} observed UGC~6728 on 6 June 2009, with the X-ray imaging spectrometer (XIS) \citep{Koyama2007} and the Hard X-ray Detector (HXD) \citep{Takahashi2007} for effective exposures of approximately $\sim46$ ks and $\sim39$ ks, respectively. The XIS observations were carried out in {\tt STANDARD} mode, whereas the HXD observation was carried out in {\tt WELL} mode. This observation was previously used by \cite{Walton2013} in their study, where they reported the presence of narrow and broad iron emission features. However, they encountered challenges in constraining the reflection parameters and consequently, fixed the inclination angle for this source at 45 degrees.

For the data reduction, we followed the standard procedures as described in the {\it Suzaku} Data Reduction Guide\footnote{\url{https://heasarc.gsfc.nasa.gov/docs/suzaku/analysis/abc/}} and applied recommended screening criteria to generate the final spectra and light curves. We used calibration files\footnote{\url{www.astro.isas.jaxa.jp/suzaku/caldb/}} (dated 2016--16--06) in {\tt FTOOLS v6.25} to reprocess the event files. Circular regions of radii 200 arcsecs and 250 arcsecs with center at the source coordinates were considered for the source and background products, respectively. The final spectrum and light curve were obtained by merging data from the two front-illuminated detectors (XIS0 and XIS3). The response files were generated using {\tt XISRESP} task. In our analysis, we excluded the data in the 1.6--2.0 keV range to avoid the effect of known Si K-edge. The XIS spectrum was subsequently binned using the {\tt GRPPHA} task, with a minimum of 100 counts per bin. For the HXD/PIN data reduction, we applied {\tt hxdpinxbpi} and {\tt hxdpinxblc} tasks on the event file to produce the spectrum and light curves. Furthermore, the HXD spectrum was corrected for the non-X-ray background, cosmic X-ray background and dead time.

\section{Result}
\label{sec:result}

\subsection{Spectral Analysis}
\label{sec:spectral}

We performed X-ray spectral analysis using data from {\it Swift}/XRT, {\it XMM-Newton}/EPIC-pn, {\it Suzaku}/XIS-HXD and {\it NuSTAR}/FPMA-FPMB instruments, covering a duration of approximately 15 years (2006-2021). The spectral analysis was carried out using $\chi^2$ statistics onto the observational data in {\tt Xspec v12.10.1f} \citep{Arnaud1996} software package. The details of the observations from multiple observatories used in the current work are presented in Table~\ref{tab:1}. For the Suzaku observation, data in the range of 0.5--30 keV are used in the spectral analysis. In 2016 and 2017, the source was observed with {\it Swift}/XRT and {\it NuSATR} simultaneously. For these observations, data in the energy ranges of 0.3--10 keV and 3--79 keV are used from XRT and NuSTAR, respectively. Apart from these simultaneous broad-band observations, UGC~6728 was also observed with {\it XMM-Newton} in 2006 and {\it Swift}/XRT in 2020 and 2021 in the soft X-ray energy band (below 10 keV).

 Data from these observations are fitted using multiple spectral models, and the error associated with each parameter of the model is determined at a 90\% confidence level (or $1.6\sigma$) using the {\tt `error'}\footnote{\url{https://heasarc.gsfc.nasa.gov/xanadu/Xspec/manual/node79.html}} task in {\tt Xspec}. The unabsorbed luminosity is calculated using the {\tt `clumin'}\footnote{\url{https://heasarc.gsfc.nasa.gov/xanadu/Xspec/manual/node286.html}} task on the composite model. The Eddington luminosity for UGC~6728 is computed as $L_{Edd}=8.91\times10^{43}$ erg/s, based on a mass of $7.1\times10^5$ $M_\odot$ \citep{Bentz2016}. In this work, we adopted the following cosmological parameters: $H_0=70 \text{~km~s}^{-1}\text{~Mpc}^{-1}; ~ \Lambda_0=0.73, \text{~and } \Omega_M=0.27$ \citep{Bennett2003}.

\subsubsection{ Model Construction}
As the X-ray spectral properties of UGC~6728 are not well understood to date, our motivation was to characterize the X-ray spectrum of this source. For this purpose, at first, we parameterize the X-ray spectrum across a broad energy range, from 0.3 to 79.0 keV, using various simple models (Power law, Gaussian, etc.). Then, we focus on the primary continuum within the energy range of 3.0 to 10.0 keV. According to current understanding, the X-ray continuum photons are produced mainly by the inverse Compton scattering of thermal photons from the accretion disc in a hot electron cloud. This process, being non-thermal, often leads to a power-law-type spectrum with a high-energy cut-off. Therefore, we begin our analysis by employing the power law ({\tt powerlaw}\footnote{\url{https://heasarc.gsfc.nasa.gov/xanadu/Xspec/manual/node216.html}}) and cut-off power law ({\tt cutoffpl}\footnote{\url{https://heasarc.gsfc.nasa.gov/xanadu/Xspec/manual/node161.html}}) models to characterize the primary continuum. The equivalent hydrogen column density along the line of sight of the source modifies the shape of the power law component. To account for this, we consider the Galactic hydrogen column density as $N_{H,gal}$ at $4.42\times10^{20}$ cm$^{-2}$\footnote{\url{https://heasarc.gsfc.nasa.gov/cgi-bin/Tools/w3nh/w3nh.pl}} using a multiplicative model {\tt TBabs}\footnote{\url{https://heasarc.gsfc.nasa.gov/xanadu/Xspec/manual/node268.html}} \citep{Verner1996} in {\tt Xspec}. Along with this, we introduce another multiplicative model component {\tt zTBabs} \citep{Wilms2000} to incorporate the extragalactic hydrogen column density ($N_{H}$) with redshift (z) $=0.0065$. So, our base model to fit the X-ray continuum is expressed as {\tt TBabs*zTBabs*constant*(powerlaw/cutoffpl)}, where {\tt constant} is used for cross-normalisation for different instruments.  

We fit each spectrum by this baseline model, and the corresponding variation of $\chi_{red}$ for the 2017 XRT2+N2 observation is shown in the top panel of Figure~\ref{fig:chi}. During the spectral fitting, a distinctive line-like feature emerged in the energy range of 6-7 keV for the SU (MJD-54988), XRT1+N1 (MJD-57579), and XRT2+N2 (MJD-58093) observations. To account this feature, we introduce a Gaussian model, {\tt zGauss}\footnote{\url{https://heasarc.gsfc.nasa.gov/xanadu/Xspec/manual/node176.html}} with z=0.0065 in {\tt Xspec} and the corresponding variation of $\chi_{red}$ for the 2017 XRT2+N2 observation is shown in the second panel of Figure~\ref{fig:chi}. 

After successfully fitting the primary continuum with the Fe-line, we extend the X-ray spectrum into the high-energy regime (above 10.0 keV) and observe no significant deviations. This suggests that the reflection component in the X-ray spectra above 10.0 keV is either absent or not significant within a 90\% confidence level. Next, we extend the X-ray spectra into the low energy regime (below 3.0 keV) and observe a significant deviation of the model from the observed spectra. Given that UGC~6728 is a Seyfert~1 AGN, it is expected to exhibit a soft-excess component below 2.0 keV. We initially model this component by {\tt powerlaw} \citep{Nandi2021,Nandi2023}. Subsequently, we fit the broadband spectra (0.3--79.0 keV) using physical models, such as {\it TCAF}, {\tt relxill}. We present the variation of $\chi_{red}$ for the 2017 XRT2+N2 observation in the last panel of Figure~\ref{fig:chi} using {\it TCAF} model fitting.

\subsubsection{ Power-law}
\label{sec:powerlaw}
We start the spectral fitting using the power-law model to characterize the primary continuum within the energy range of 3.0-10.0 keV. This model, represented in {\tt Xspec} as {\tt tbabs*zTbabs*(powerlaw+zGauss)}, is used as the baseline for our analysis. The {\tt zGauss} component is used to compensate for the Fe-line at 6.4 keV. Later, we substitute the power law model with a cut-off power law model, represented as {\tt cutoffpl}. The spectral fitting results are presented in Table~\ref{tab:cutpoffpl}. It is to be noted that there are no significant variations in the spectral index ($\Gamma$) observed in either case.

For the spectral analysis of XMM-Newton (MJD--53789) data in the 3.0-10.0 keV range, we employ {\tt tbabs*tbabs*cutoffpl} model in our fitting. This model provided a satisfactory fit with a spectral index of $\Gamma=1.39^{+0.10}_{-0.11}$ and cut-off energy of $E_c=489.54^{+258.54}_{-254.62}$ keV, resulting in a $\chi^2$ value of 42.71 for 37 degrees of freedom (dof). The continuum luminosity in the 2.0--10.0 keV range is calculated as $\log(L_x)=41.85\pm0.28$. No Fe-line was detected during this observation. 

Applying a similar methodology, we analyze 0.5-30.0 keV data from the Suzaku (MJD--54988) observation of the source. The estimated spectral index ($\Gamma$) and cut-off energy ($E_c$) are determined to be $1.50^{+0.06}_{-0.07}$ and $469.65^{+155.52}_{-271.71}$ keV, respectively. During this analysis, we also detect an Fe-line at $6.54^{+0.10}_{-0.09}$ keV with an equivalent width ($EW$) of $188^{+15}_{-17}$ eV. The normalization ($N$) is calculated to be $12.2^{+2.07}_{-1.88}\times10^{-3}$ photons/keV/cm$^{-2}$/s, resulting in a reduced $\chi^2$ ($\chi^2_{red}$) of 1.02. Along with this, the 2.0-10.0 keV source luminosity is determined to be $\log(L_x)=43.03\pm0.45$.

We proceed to analyze the XRT1+N1 (MJD--57579) observation using the baseline model. During this analysis, we observe changes in the spectral parameters. The spectral index ($\Gamma$) changed from $1.39^{+0.10}_{-0.11}$ to $1.64^{+0.04}_{-0.04}$, and the corresponding cut-off energy ($E_c$) varied from $389.54^{+258.54}_{-254.62}$ to $376.45^{+287.82}_{-266.78}$ keV. The normalization ($N$) is determined to be $1.30^{+0.16}_{-0.14}\times10^{-3}$ photons/keV/cm$^{-2}$/s, and the reduced $\chi^2$ ($\chi^2_{red}$) value for this fitting is 1.08 for 162 $dof$. During the spectral fitting, we detect an Fe-line at $6.40^{+0.02}_{-0.02}$ keV with an equivalent width ($EW$) of $104^{+8}_{-9}$ eV. The result is listed in Table~\ref{tab:cutpoffpl}. The continuum luminosity in the 2.0 to 10.0 keV range is estimated to be $\log(L_x)=41.83\pm0.35$.

In the subsequent analysis, we employ the combined observations from XRT2+N2 (MJD--58093) to conduct spectral analysis using the cut-off power-law model. The results indicate a spectral index ($\Gamma$) of $1.69^{+0.03}_{-0.02}$ and a cut-off energy ($E_c$) of $344.89^{+100.15}_{-157.83}$ keV. The goodness of fit is evaluated by a $\chi^2$ value of 685.54 for 665 $dof$. During this analysis, an Fe-line was detected at $6.41^{+0.01}_{-0.01}$ keV with equivalent width $EW=80^{+7}_{-8}$ eV and the result is listed in Table~\ref{tab:cutpoffpl}. The continuum luminosity in the energy range of 2.0--10.0 keV is estimated to be  $\log(L_x)=42.14\pm0.37$. 

Using the same model for the remaining {\it Swift}/XRT observations, XRT3 (MJD--58097) and XRT4 (MJD--59414), we determine the photon indices as $\Gamma=1.66^{+0.11}_{-0.10}$ and $\Gamma=1.85^{+0.06}_{-0.07}$, with cut-off energies of $E_c=359.60^{+195.78}_{-168.74}$ keV and $E_c=258.60^{+101.88}_{-129.26}$ keV, respectively.  The calculated luminosity within the 2.0-10.0 keV energy range is $\log(L_x)=41.78\pm0.13$ and $\log(L_x)=41.90\pm0.35$, respectively.

\begin{table*}
	  \caption{\label{tab:cutpoffpl} {\tt cutoffpl} fitting result for all spectra. The unabsorbed luminosity is calculated for the energy range from 2.0 to 10.0 keV.}	
   \small
  	{\centerline{}}
  	\begin{center}
  		\begin{tabular}{c c c c c c c c c c c }
  			\hline
\rm ID & \rm MJD  &  \rm $N_H$          & \rm $\Gamma$& \rm $E_c$ & \rm Fe $K_\alpha$ & \rm EW & \rm $N_{Fe}$  & \rm $N$     &  \rm $\chi^2/dof$ & \rm $\log(L)$\\
       &          &  $(10^{20})$        &             &   (keV)   &  &  & $(10^{-6})$  & $(10^{-3})$ &  &  \\
\hline
&&&&&& \\
 XMM  &  53789    & $0.10^{+0.15}_{-0.12}$ &$1.39^{+0.10}_{-0.11}$ &$489.54^{+287.82}_{-254.62}$ & -- & -- & -- & $1.10^{+0.07}_{-0.08}$ & $42.71/37$ & $41.85\pm0.28$ \\
 &&&&&& \\
 SU  &  54988    & $0.15^{+0.11}_{-0.11}$ &$1.50^{+0.06}_{-0.07}$ &$469.65^{+155.52}_{-171.71}$ & $6.54^{+0.10}_{-0.11}$ & $188^{+15}_{-17}$ & $361.4^{+3.84}_{-4.57}$ &  $12.2^{+2.07}_{-1.88}$ & $327.31/320$ & $43.03\pm0.45$ \\
 &&&&&& \\
 XRT1+N1 & 57579  & $0.10^{+0.10}_{-0.10}$ &$1.64^{+0.04}_{-0.04}$ &$376.45^{+258.54}_{-266.78}$ & $6.40^{+0.02}_{-0.02}$ & $104^{+8}_{-9}$ & $6.51^{+0.15}_{-0.15}$ & $1.30^{+0.16}_{-0.14}$ & $175.68/162$ & $41.83\pm0.35$  \\
 &&&&&& \\
 XRT2+N2 & 58093  & $0.10^{+0.12}_{-0.10}$ &$1.69^{+0.03}_{-0.02}$ &$344.89^{+100.15}_{-157.83}$ & $6.41^{+0.01}_{-0.01}$ & $80^{+7}_{-8}$ & $11.6^{+1.11}_{-1.10}$ &  $3.31^{+0.22}_{-0.23}$ & $685.54/665$ & $42.14\pm0.37$  \\
 &&&&&& \\
XRT3 & 58097  & $0.09^{+0.13}_{-0.09}$ &$1.66^{+0.11}_{-0.10}$ &$359.60^{+195.78}_{-168.74}$ & -- & -- & --  & $1.46^{+0.26}_{-0.24}$ & $14.31/17$ & $41.78\pm0.33$  \\
 &&&&&& \\
 XRT4 & 59414  & $0.11^{+0.15}_{-0.12}$ &$1.85^{+0.06}_{-0.07}$ &$258.60^{+101.88}_{-129.26}$ & -- & -- & --  & $2.70^{+0.27}_{-0.27}$ & $22.81/27$ & $41.90\pm0.35$ \\
 &&&&&& \\
 \hline
 \end{tabular}
 \end{center}
 \end{table*}


\subsubsection{Nthcomp}
\label{sec:Nthcomp}

\begin{figure} 
	\centering
	\includegraphics[scale=0.7]{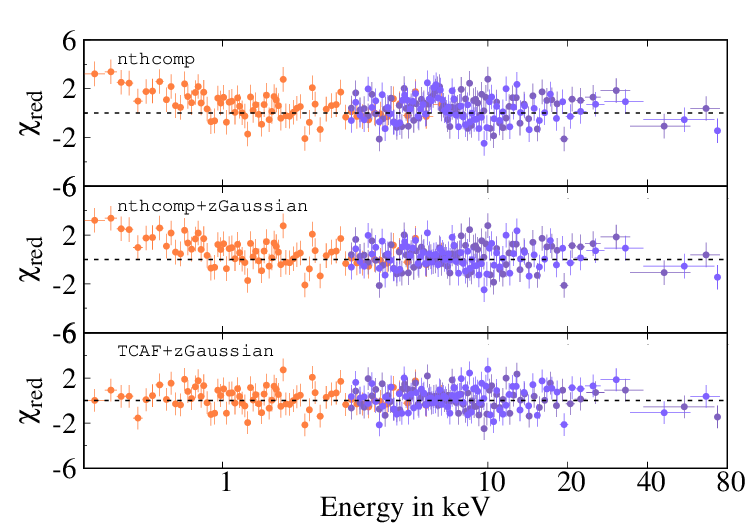}
	\caption{Variation of $\chi_{red}$ is shown for each model component while fitting the broadband spectrum of UGC~6728 during the 2017 XRT2+N2 observation. Initially, we fitted the data above $3.0$ keV with the {\tt Nthcomp} model. Then we added {\tt zGaussian} for the Fe-line at $\sim6.41$ keV. Later, we replaced {\tt Nthcomp} by {\tt TCAF} for the broadband spectrum ($0.3-79.0$ keV).}
	\label{fig:chi} 
\end{figure}

In order to understand the physical processes leading to the primary continuum X-ray emission from UGC~6728, we fit the broadband energy spectrum with the thermal Comptonization model, {\tt Nthcomp}\footnote{\url{https://heasarc.gsfc.nasa.gov/xanadu/Xspec/manual/node205.html}} \citep{Zdziarski1996, Zycki1999}. The model depends on the seed photon energy $(kT_{bb})$, which we consider to be $200$ eV, corresponding to the Wien's temperature of the disc for this source. It is worth mentioning that we tested a range of values of $(kT_{bb})$ from $150$ to $250$ eV and found no significant deviations in the spectral fitting. For our analysis, we opted $inp\_type=1$, which is the case when the seed photons are of a multicolour blackbody nature. The composite model we used in {\tt Xspec} reads as
\begin{center}
		{\tt TBabs*zTBabs*constant*(Nthcomp+zGauss)}.
\end{center}
For the spectral fitting, we initially focus on the $3.0$ to $10.0$ keV range spectra and apply the asymptotic power law model. Subsequently, we extend the spectra above $10.0$ keV and below $3.0$ keV, keeping the photon index constant. Then, we apply the {\tt Nthcomp} model replacing the {\tt power-law} component in the broadband spectral fitting. Interestingly, we observe positive residuals below $2.0$ keV, indicating the presence of soft-excess emission in this source. Additionally, we provide annotations for the model components of the broadband spectral fitting ($0.3-79.0$ keV range) during the 2017 (XRT2+N2) observation in the respective panels. We illustrate the variation of $\chi^2$ for the {\tt Nthcomp} model in Figure~\ref{fig:chi}. 

We begin the spectral fitting using spectra obtained from the 2006 {\it XMM-Newton} (XMM) observation. For the continuum spectrum (above $3.0$ keV), we find the spectral index  $(\Gamma)=1.38\pm0.14$. We replace the {\tt powerlaw} model with {\tt Nthcomp} for the primary continuum by fixing $(\Gamma)$ at $1.38$. The estimated electron temperature $(kT_e)$ from the spectral fitting is $475.89\pm250.35$ keV, accompanied by a {\tt Nthcomp} normalization of $1.10^{+0.07}_{-0.08}\times 10^{-3}$ photons/keV/sec/cm$^2$. The reduced $\chi^2$ ($\chi^2_{red}=\chi^2/\text{dof}$) is found to be $1.08$ for 37 degrees of freedom (dof). 

Using the baseline model, we proceed to fit data from the Suzaku observation (MJD--54988), and the corresponding results are presented in Table~\ref{tab:spec_fit_Nthcomp}. The spectral fitting provides the photon index, $\Gamma$, of $1.50\pm0.04$ and an electron energy, $kT_e$, of $350.42\pm151.74$ keV. Using the values of $\Gamma$ and $kT_e$ in Equation~\ref{eq:1}, the value of the optical depth, $\tau$, is determined to be $0.68^{+0.71}_{-0.34}$. A Fe-line at $6.54\pm0.10$ with an equivalent width (EW) of $188^{+15}_{-17}$ eV is detected in the source spectrum obtained from this observation. The normalization is found to be $7.65^{+1.07}_{-2.0}$ photons/keV/sec/cm$^2$ with a reduced $\chi^2/dof=277.31/319$.


We employ similar spectral fitting procedures for the 2016 (XRT1+N1) and 2017 (XRT2+N2) observations. The estimated values of $\Gamma$ are $1.64\pm0.03$ and $1.69\pm0.02$ for the 2016 and 2017 observations, respectively. The electron temperatures estimated from the X-ray broadband ($0.3$-$79.0$ keV range) spectral fitting remain at $256.40\pm86.21$ keV and $249.18\pm41.95$ keV, respectively (see Table~\ref{tab:spec_fit_Nthcomp}). Additionally, We identify the presence of an iron K$_\alpha$ line at $6.40\pm0.02$ and $6.41\pm0.01$ keV for XRT1+N1 and XRT2+N2 observations with equivalent widths (EW) of $104^{+3}_{-4}$ and $80.9^{+3}_{-3}$ eV, respectively. The normalizations of the {\tt Nthcomp} model fitting of these broadband X-ray spectra are $1.27^{+0.16}_{-0.14}\times 10^{-3}$ and $3.23^{+0.22}_{-0.23}\times 10^{-3}$ photons/keV/sec/cm$^2$, respectively. The corresponding $\chi^2_{red}$ values are $1.08$ and $1.03$ with $161$ and $662$ dofs (see Table~\ref{tab:spec_fit_Nthcomp}), respectively. 

Subsequently, we carry out spectral analysis of the {\it Swift}/XRT observations from 2020 (XRT3) and 2021 (XRT4) using {\tt Nthcomp} model. No iron lines are detected in the spectra from these observations. The power-law indices $(\Gamma)$ from the spectral fitting are $1.66\pm0.10$ and $1.85\pm0.05$ for XRT3 and XRT4, respectively and the corresponding electron temperatures $(kT_e)$ are $248.78\pm124.36$ and $178.32\pm126.32$ keV, respectively. 

\begin{figure} 
	\centering
	\includegraphics[scale=0.7]{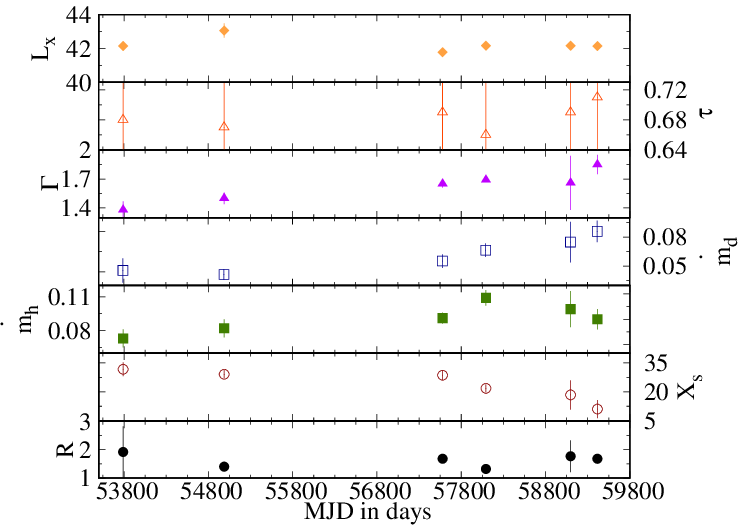}
	\caption{Variation of different parameters of {\tt powerlaw} fitting with time for all spectra. The corresponding parameters are given in Table~\ref{tab:spec_fit_Nthcomp}.}
	\label{fig:parameter} 
\end{figure}

Through the comprehensive spectral fitting of X-ray broadband data from all observations using the {\tt Nthcomp} model, we determine that the hydrogen column density $(N_H)$ along the line of sight is negligible and remains relatively constant throughout our observations. This parameter, $(N_H)$ varies from $0.09^{+0.13}_{-0.09}\times10^{20}$ cm$^{-2}$ to $0.12^{+0.15}_{-0.12}\times10^{20}$ cm$^{-2}$, where the Galactic hydrogen column density $(N_{H,gal})$ is fixed at $4.42\times10^{20}$ cm$^{-2}$. Additionally, we observe an increase in the photon index $(\Gamma)$ from $1.38\pm0.4$ to $1.85\pm0.5$, while the electron temperature $(kT_e)$ exhibited a decrease from $475.89\pm250.35$ keV to $178.32\pm126.32$ keV. Furthermore, we calculate the optical depth $(\tau)$ for each observation using the formula \citep{Zdziarski1996}:
\begin{equation}
\label{eq:1}
    \tau=\sqrt{\frac{9}{4}+\frac{3}{\theta_e(\Gamma+2)(\Gamma-1)}}-\frac{3}{2}.
\end{equation}
Here, $\theta_e=\frac{kT_e}{m_ec^2}$ is the thermal energy of electrons with respect to the rest mass energy. The values of $\tau$ for each observation are given in Table~\ref{tab:spec_fit_Nthcomp}. The maximum possible error in $\tau$ is calculated using the formula $\delta\tau\sim(\frac{1}{2}\frac{\Delta\theta_e}{\theta_e}+\frac{\Delta\Gamma}{\Gamma})\times\tau$, where $\Delta\theta_e$ and $\Delta\Gamma$ are taken from the estimated errors presented in Table~\ref{tab:spec_fit_Nthcomp}. The variation of $\tau$ is presented in Table~\ref{tab:spec_fit_Nthcomp} as well as in Figure~\ref{fig:parameter} (second panel from top). We find the values of optical depth to be always below $1.0$. This suggests that the Compton cloud of the source was optically thin for the entire observation period. 

  \begin{table*}
	  \caption{\label{tab:spec_fit_Nthcomp} Results obtained from the spectral fitting above 3.0 keV spectra with {\tt Nthcomp} model. The optical depth $\tau$ is calculated using Equation~\ref{eq:1}.} 	
  	\vskip -0.2 cm
  	{\centerline{}}
  	\begin{center}
  		\begin{tabular}{c c c c c c c c }
  			\hline
\rm ID & \rm MJD  &  \rm $N_H$          & \rm $\Gamma$  & \rm $kT_e$     & \rm $N$       & \rm $\chi^2/dof$ & \rm $\tau^\dagger$ \\
       &          &  $(10^{20})$        &        &  (keV)     & $(10^{-3})$ &   &   \\
\hline
&&&&& \\
 XMM  &  53789    & $0.12^{+0.15}_{-0.12}$ &$1.38^{+0.14}_{-0.14}$& $475.89\pm250.35$ & $1.10^{+0.07}_{-0.08}$  & $40.02/37$& $0.68^{+0.69}_{-0.26}$  \\
 &&&&& \\
  SU  &  54988    & $0.15^{+0.11}_{-0.12}$ &$1.50^{+0.04}_{-0.04}$& $350.42\pm151.74$ & $7.65^{+1.07}_{-2.0}$ &  $277.31/319$& $0.68^{+0.71}_{-0.34}$  \\
 &&&&& \\
 XRT1+N1 & 57579  & $0.10^{+0.10}_{-0.10}$ &$1.64^{+0.03}_{-0.03}$ & $256.40\pm86.21$ & $1.27^{+0.16}_{-0.14}$ & $174.10/161$& $0.69^{+0.62}_{-0.24}$ \\
 &&&&& \\
 XRT2+N2 & 58093  & $0.10^{+0.12}_{-0.10}$ &$1.69^{+0.02}_{-0.02}$ & $249.18\pm41.95$ & $3.23^{+0.22}_{-0.23}$ & $680.27/662$ & $0.66^{+0.52}_{-0.22}$  \\
 &&&&& \\
 XRT3 & 59097  & $0.09^{+0.13}_{-0.09}$ &$1.66^{+0.10}_{-0.10}$ & $248.74\pm124.36$ &$1.45^{+0.26}_{-0.24}$ & $14.20/14$& $0.69^{+0.51}_{-0.28}$  \\
 &&&&& \\
 XRT4 & 59414  & $0.12^{+0.15}_{-0.12}$ &$1.85^{+0.05}_{-0.05}$ & $178.32\pm126.32$ & $2.72^{+0.27}_{-0.27}$ & $21.77/26$& $0.71^{+0.55}_{-0.32}$  \\
 &&&&& \\
 \hline
 \end{tabular}
 \end{center}
 \end{table*}

\subsubsection{CompTT}
\label{sec:CompTT}

From the thermal Comptonization model ({\tt Nthcomp}), we derive the temperature of the Compton cloud and spectral index from the model fitting. We also estimate the optical depth using Equation~\ref{eq:1}. However, it is important to note that for most of the observations covering energies up to 10.0 keV, constraining the electron temperature from spectral fitting is challenging, resulting in relatively large errors in optical depth calculations. To validate the derived values of these fundamental parameters, we employ another physical model called {\tt CompTT}\footnote{\url{https://heasarc.gsfc.nasa.gov/xanadu/Xspec/manual/node158.html}} on the X-ray spectra of UGC~6728. The {\tt CompTT} \citep{ST1980, Titarchuk1994, Hua1995, TL1995} model is characterized by the input soft photons with a temperature ($T_0$) determined from the Wien's law. This model incorporates the plasma temperature of the Compton cloud ($kT_e$), optical depth ($\tau$), and the assumed geometry (disc, sphere, analytical approximation). In our analysis, we adopt a spherical geometry. The default input value for the soft photon temperature is set at 200 eV (0.2 keV) in the fitting. It is to be noted that the parameter $T_0$ is treated as a free parameter in the fitting. The composite model we used in {\tt Xspec} reads as,
\begin{center}
		{\tt TBabs*zTBabs*constant(CompTT+zGauss)}.
\end{center}

We begin our spectral fitting of data above 3.0 keV from the XMM-Newton observation in 2006 using this model, and the corresponding results are presented in Table~\ref{tab:CompTT}. For this observation, we find the soft photon temperature $(T_0)$ at $0.21\pm0.14$ keV with Compton cloud temperature $(kT_e)$ at $470.54^{+145.62}_{-148.26}$ keV. The optical depth $(\tau)$ for spherical cloud is estimated as $0.75^{+0.42}_{-0.35}$ and the model normalization is found to be $8.95^{+2.85}_{-0.86}\times10^{-6}$ photons/keV/sec/cm$^{2}$. During the model fitting, we did not observe any significant deviations of the observational data from the continuum model around $\sim6.4$ keV. Therefore, we discard the model component {\tt zGauss} for this particular spectral fitting of {\it XMM-Newton} data.

We follow a similar methodology for fitting the data from the Suzaku observation (MJD--54988) within the energy range of 3.0 to 30.0 keV. As with the {\tt Nthcomp} model fitting (see \ref{sec:Nthcomp}), we initially consider the 3.0 keV to 10.0 keV spectrum for this observation when employing the {\tt CompTT} model. Then, we extend the observed spectrum to higher-energy regions and perform fitting accordingly. However, we are unable to distinguish the reflection component above 10.0 keV due to poor data quality. It is plausible that the reflection was either absent or barely present during this observation. While conducting the spectral fitting, we identify the presence of Fe-line at $6.54^{+0.10}_{-0.09}$ keV with an equivalent width of $290^{+15}_{-17}$ eV. From the continuum fitting, using the composite model described above, we determine the following parameters: soft photon temperature $T_0=0.20^{+0.15}_{-0.18}$ keV, temperature of the Compton cloud $kT_e=386.87^{+103.08}_{-098.65}$ keV and optical depth $\tau=0.61^{+0.32}_{-0.31}$.

We then attempt to fit two broadband X-ray spectra from 2016 and 2017 observations of UGC~6728 obtained from the simultaneous {\it Swift}/XRT and {\it NuSTAR} observations, covering the energy band of 3.0 to 79.0 keV. In both the observations, we detect a narrow Fe-line at $6.4\pm0.02$ keV with an equivalent width of $104^{+8}_{-9}$ eV and $80^{+7}_{-8}$ eV for XRT1+N1 (MJD--57579) and XRT2+N2 (MJD--58093) observations, respectively. The disc temperature ($T_0$) is found to be consistent for both observations at $0.18\pm0.13$ keV. The temperature of the Compton cloud ($kT_e$) is found to be $215.22^{+59.36}_{-58.74}$ keV for XRT1+N1 and $251.41^{+77.41}_{-82.26}$ keV for XRT2+N2. The optical depth vary between $0.87^{+0.21}_{-0.23}$ and $0.62^{+0.35}_{-0.41}$ for these two observations. The model normalizations for these observations are estimated to be $34.8^{+8.14}_{-9.54}\times10^{-6}$ photons/keV/sec/cm$^{2}$ for XRT1+N1 and $16.0^{+3.76}_{-4.19}\times10^{-6}$ photons/keV/sec/cm$^{2}$ for XRT2+N2. Detailed results are provided in Table~\ref{tab:CompTT}. 

In the subsequent analysis, we perform spectral fitting using {\it Swift}/XRT data. The disc temperature ($T_0$) is found to be consistent at $0.14^{+0.17}_{-0.19}$ for the observations XRT3 and XRT4. The temperature of the Compton cloud ($kT_e$) is found to be $248.78^{+126.35}_{-146.24}$ keV and $215.06^{+104.63}_{-151.21}$ keV for XRT3 and XRT4, respectively. The optical depth is found to vary accordingly. The estimated values of $\tau$ are found to be $0.74^{+0.32}_{-0.24}$ for XRT3 and $0.39^{+0.41}_{-0.28}$ for XRT4. The Fe-line is found to be absent in these two observations, leading us to exclude the {\tt zGauss} component from the composite model. The normalizations for the continuum of these observations using the {\tt CompTT} model are estimated to be $16.0^{+3.76}_{-4.19}\times10^{-6}$ photons/keV/sec/cm$^{2}$ for XRT3 and $52.1^{+7.47}_{-8.15}\times10^{-6}$ photons/keV/sec/cm$^{2}$ for XRT4. The parameters obtained from the {\tt CompTT} model fitting are presented in Table~\ref{tab:CompTT}. 

Throughout the long-term (2006--2021) X-ray spectral fitting of the continuum of UGC~6728 using {\tt CompTT} model, we find that the extra-galactic hydrogen column density $(N_H)$ was nearly constant and insignificant. This is attributed to the `bare' nature of the nucleus of this AGN. The disc temperature $(T_0)$ vary between $0.14\pm0.2$ to $0.21\pm0.14$ keV. The Compton cloud temperature $(kT_e)$ and the optical depth $(\tau)$ also exhibit variations between $215.06^{+104.63}_{-151.21}$ to $470.54^{+145.62}_{-148.26}$ keV and $0.87^{+0.21}_{-0.23}$ to $0.39^{+0.41}_{-0.28}$, respectively. These variations of $kT_e$ and $\tau$ closely resemble that observed from the thermal Comptonization model ({\tt Nthcomp}) fitting. The optical depth $\tau$ is found to be consistently below 1.0, suggesting that the cloud was optically thin throughout the observational period.

  \begin{table*}
	  \caption{\label{tab:CompTT} Results obtained from the spectral fitting of 3.0--10.0 keV data with {\tt CompTT} model.} 	
  	\vskip -0.2 cm
  	{\centerline{}}
  	\begin{center}
  		\begin{tabular}{c c c c c c c c}
  			\hline
\rm ID & \rm MJD  &  \rm $N_H$          & \rm $T_0$  & \rm $kT_e$ & \rm $\tau$     & \rm $N$    & \rm $\chi^2/dof$  \\
       &          &  $(10^{20})$        &   (keV)    &  (keV)     &                & $(10^{-6})$&                   \\
\hline
&&&&& \\
 XMM  &  53789    & $0.12^{+0.12}_{-0.13}$ &$0.21^{+0.14}_{-0.14}$& $470.54^{+145.62}_{-148.26}$ & $0.75^{+0.42}_{-0.35}$ & $8.95^{+2.85}_{-0.86}$  & $39.71/32$  \\
&&&&&& \\
 SU  &  54988    & $0.15^{+0.22}_{-0.11}$ &$0.20^{+0.15}_{-0.18}$& $386.87^{+103.08}_{-098.65}$ & $0.61^{+0.32}_{-0.31}$ & $44.1^{+6.51}_{-7.17}$  & $197.36/189$  \\
&&&&&& \\
 XRT1+N1&57579    & $0.11^{+0.12}_{-0.12}$ &$0.18^{+0.12}_{-0.13}$& $215.22^{+59.36}_{-58.74}$ & $0.87^{+0.21}_{-0.23}$ & $19.9^{+7.04}_{-10.2}$  & $173.31/159$  \\
&&&&&& \\
 XRT2+N2&58093    & $0.10^{+0.11}_{-0.11}$ &$0.19^{+0.13}_{-0.15}$& $251.41^{+77.41}_{-82.26}$ & $0.62^{+0.35}_{-0.41}$ & $34.8^{+8.14}_{-9.54}$  & $683.22/661$  \\
&&&&&& \\
 XRT3   &59097    & $0.09^{+0.12}_{-0.09}$ &$0.15^{+0.15}_{-0.14}$& $248.78^{+126.35}_{-146.24}$ & $0.74^{+0.32}_{-0.24}$ & $16.0^{+3.76}_{-4.19}$  & $26.01/24$  \\
&&&&&& \\
 XRT4   &59414    & $0.12^{+0.10}_{-0.11}$ &$0.14^{+0.17}_{-0.19}$& $215.06^{+104.63}_{-151.21}$ & $0.39^{+0.41}_{-0.28}$ & $52.1^{+7.47}_{-8.15}$  & $51.40/54$  \\

 \hline
 \end{tabular}
 \end{center}
 \end{table*}

\subsubsection{TCAF}

By conducting {\tt Nthcomp} model fitting of X-ray spectra spanning over approximately 15 years $(2006-2021)$, we derive several valuable physical parameters related to the spectral hardness and electron temperature of the Compton cloud or corona in UGC~6728. The calculated value of optical depth consistently indicated an optically thin Compton cloud. Nevertheless, fundamental properties such as the mass of the central object, accretion flow rates, and size of the Compton cloud for this accreting supermassive black hole are still not derived from the above analysis. Knowledge of these parameters would contribute to a deeper comprehension of this massive accreting system. To address this, we employ the Two Component Advective Flow (TCAF) model \citep{CT1995} to analyze the X-ray spectrum of this object. For the spectral fitting, the model used is

\begin{center}
		{\tt TBabs*zTBabs*constant(TCAF+zGauss)}.
\end{center}
The TCAF model is based on four accretion flow parameters:  (i) Keplerian disc accretion rate ($\dot{m}_d$) in units of the Eddington rate ($\Dot{m}_{Edd}$), (ii) Sub-Keplerian halo accretion rate ($\dot{m}_h$) in units of the Eddington rate ($\Dot{m}_{Edd}$), (iii) shock compression ratio (R), and (iv) shock location ($X_s$) in units of the Schwarzschild radius ($r_g=2GM/c^2$) with an intrinsic parameter which is the mass of the central black hole in units of the solar mass ($M_\odot$). The upper limit, lower limit, and default values of each parameter are kept in a file called {\it lmodel.dat}. These values are provided in Table~\ref{tab:tcaf_parameter_space}. We use {\tt initpackage} and {\tt lmod} task in {\tt Xspec} to run this table to fit the X-ray spectrum of UGC~6728 with the TCAF model. We run the source code about $\sim10^5$ times and select the best spectrum by minimizing $\chi$.

  \begin{table*}
	  \caption{\label{tab:tcaf_parameter_space} The TCAF parameter space is defined in the file lmod.dat. The two columns for minima and maxima are provided for the range of iterations. Between them, the 1st column indicates the soft bound and the 2nd column gives the hardbound of the parameters.}	
  	\vskip -0.2 cm
  	{\centerline{}}
  	\begin{center}
  		\begin{tabular}{c c c c c c c c}
  			\hline
\rm Model parameters & \rm Parameter units  &  \rm Default value  & \rm Min.  & Min.  &  Max.  &  Max.  & Increment \\
  			\hline
$\rm M_{BH}$&$\rm M_\odot$&$\rm 1.0\times10^5$&$2\times10^4$&$2\times10^4$&$5.5\times10^8$&$5.5\times10^8$&$ 10.0 $ \\
$\dot{m}_d $& $\rm \Dot{m}_{Edd} $ & $ \rm 0.001 $ & $ 0.0001 $ & $ 0.0001 $ & $ 1.0 $ & $ 2.0 $ & $ 0.0001 $ \\
$\dot{m}_h $& $\rm \Dot{m}_{Edd} $ & $ \rm 0.01 $ & $ 0.0001 $ & $ 0.0001 $ & $ 2.0 $ & $ 3.0 $ & $ 0.0001 $ \\
$\rm X_s $& $\rm r_g $ & $ \rm 50.0 $ & $ 8.0 $ & $ 8.0 $ & $ 1000.0 $ & $ 1000.0 $ & $ 1.0 $ \\
$\rm R $& $\rm  $ & $ \rm 2.5 $ & $ 1.01 $ & $ 1.01 $ & $ 6.8 $ & $ 6.8 $ & $ 0.1 $ \\
\hline
\end{tabular}
\end{center}
\end{table*}

\begin{figure*}
		\centering
  \includegraphics[width=0.29\textwidth, angle =0]{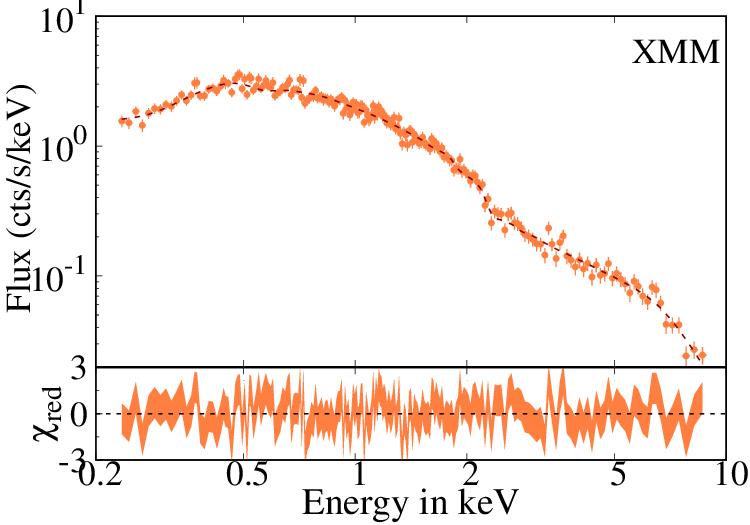}
  \includegraphics[width=0.29\textwidth, angle =0]{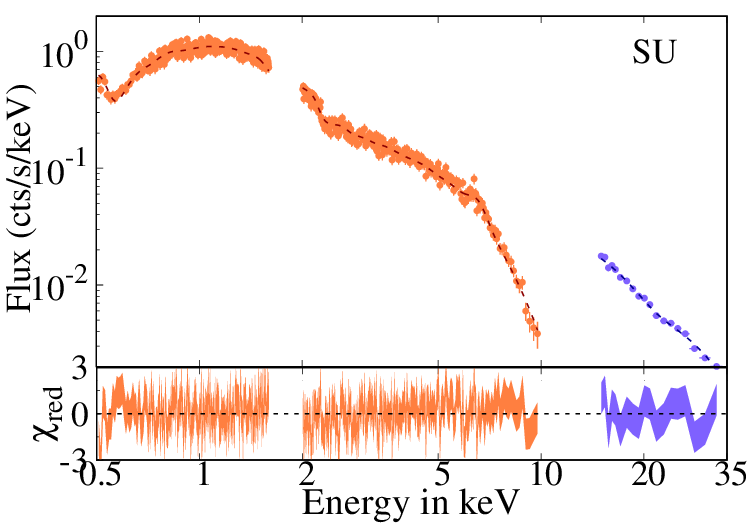}
  \includegraphics[width=0.29\textwidth, angle =0]{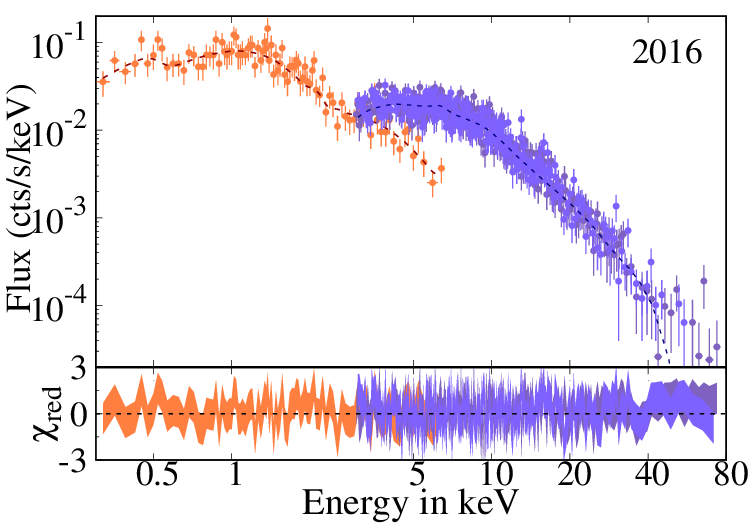}
  \includegraphics[width=0.29\textwidth, angle =0]{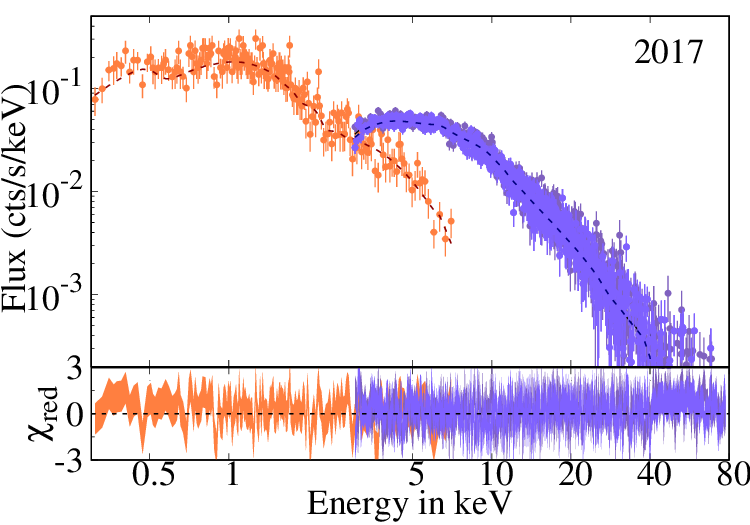}
  \includegraphics[width=0.29\textwidth, angle =0]{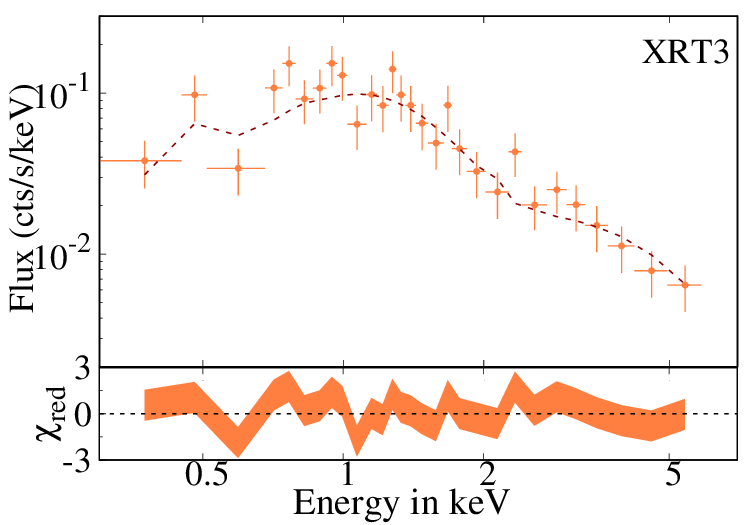}
  \includegraphics[width=0.29\textwidth, angle =0]{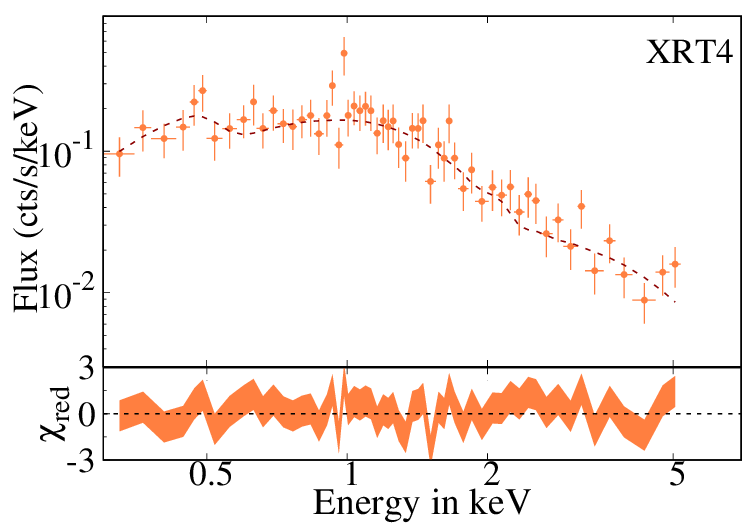}
		\caption{ {\tt TCAF} model fitted X-ray spectra of UGC~6728 from various observations from 2009-2021, along with the residuals obtained from the spectral fitting.}
		\label{fig:all_spec}
	\end{figure*}

We begin the spectral fitting process of X-ray data from the XMM-Newton observation (2006) using the baseline model as described above. During model fitting, any significant deviation in spectral fit near $\sim6.4$ keV is not seen. Therefore, the Gaussian component {\tt zGauss} is excluded from the fitting procedure. The fitted parameters obtained from the {\tt TCAF} model for this specific observation are presented in Table~\ref{tab:spec_fit_TCAF} and the model fitted spectrum with the variation of $\chi$ is presented in Figure~\ref{fig:all_spec}. For this observation, we obtain $M_{BH}=7.28^{+1.09}_{-1.08}\times10^5$ M$_\odot$ for the central SMBH with disc and halo mass accretion rates $(\Dot{m}_{d})$ and $(\Dot{m}_{h})$, of $0.05^{+0.01}_{-0.01}$ $\Dot{m}_{Edd}$ and $0.07^{+0.01}_{-0.01}$ $\Dot{m}_{Edd}$, respectively. Furthermore, we determine the shock location or the boundary of the Compton cloud $(X_s)$ to be at $31.58^{+4.65}_{-3.56}$ $r_g$ with a shock strength of $(R)=1.92^{+0.99}_{-0.91}$ for this XMM-Newton observation. 

Using a similar approach, we fit the broadband ($0.5-30.0$ keV) X-ray spectra from Suzaku (MJD--54988), and the corresponding result is shown in Table~\ref{tab:tcaf_parameter_space} and the spectra are presented in Figure~\ref{fig:all_spec}. The estimated values of model parameters are : $M_{BH}=7.09^{+0.59}_{-0.57}\times10^5$ M$_\odot$, $\Dot{m}_d=0.05^{+0.01}_{-0.01}$ $\Dot{m}_{Edd}$, $\Dot{m}_h=0.08^{+0.01}_{-0.01}$ $\Dot{m}_{Edd}$, $X_s=29.09^{+2.10}_{-2.09}$ $r_g$, and $R=1.40^{+0.07}_{-0.09}$. The normalization of the continuum model is estimated as $335.4^{+5.65}_{-8.62}\times10^{-5}$ photons/keV/sec/cm$^{2}$ and $\chi^2/dof=277.31/320$. The X-ray luminosity for the energy band 0.5--10.0 keV is found to be $\log(L_x) = 43.06\pm0.43$. 

Next, we analyze the broadband ($0.3-79.0$ keV) X-ray spectra from XRT1+N1 (2016) and XRT2+N2 (2017) using the baseline model mentioned earlier. Our analysis yielded the following estimation of the {\tt TCAF} model parameters for 2016 observation as : $M_{BH}=7.11^{+0.39}_{-0.32}\times10^5$ M$_\odot$, $\Dot{m}_d=0.06^{+0.01}_{-0.01}$ $\Dot{m}_{Edd}$, $\Dot{m}_h=0.09^{+0.01}_{-0.01}$ $\Dot{m}_{Edd}$, $X_s=23.62^{+2.55}_{-2.56}$ $r_g$, and $R=1.68^{+0.09}_{-0.08}$. Similarly, for the 2017 observation, the corresponding parameter values are estimated as $M_{BH}=7.14^{+0.22}_{-0.19}\times10^5$ M$_\odot$, $\Dot{m}_d=0.07^{+0.01}_{-0.01}$ $\Dot{m}_{Edd}$, $\Dot{m}_h=0.11^{+0.01}_{-0.01}$ $\Dot{m}_{Edd}$, $X_s=21.13^{+2.07}_{-1.56}$ $r_g$, and $R=1.34^{+0.06}_{-0.05}$. We observe that the normalization for XRT1+N1 (2016) observation is $1.51^{+0.37}_{-0.38}\times10^{-5}$ photons/keV/sec/cm$^{2}$, which is increased to $1.81^{+0.80}_{-0.78}\times10^{-5}$ photons/keV/sec/cm$^{2}$ during the XRT2+N2 (2017) observation. We observe a change in X-ray luminosity from $Lx \sim 6.0\times10^{41}$ (XRT1+N1: 2016) to $Lx \sim 1.5\times10^{42}$ erg/s (XRT2+N2: 2017) in about 500 days. It is to be noted that this type of spectral variability in a short timescale is commonly seen in AGNs \citep{Ballantyne2014, Zoghbi2017}. The details of the parameter space are presented in Table~\ref{tab:spec_fit_TCAF}, and corresponding {\tt TCAF} fitted spectra from these two observations are shown in Figure~\ref{fig:all_spec} along with the variation of $\chi$. It is to be noted that, for these broadband spectra, we find the Fe~K$_\alpha$ line at $6.4\pm0.02$ and $6.41\pm0.01$ keV for XRT1+N1 and XRT2+N2 observations with equivalent width (EW) of $104^{+8}_{-9}$ and $80.9^{+7}_{-8}$ eV, respectively. 

We then proceed to fit the data from the last two {\it Swift}/XRT observations, XRT3 and XRT4, using the baseline model. No Fe-line is observed in these spectra. From the {\tt TCAF} model fitting, we estimate the mass of the central black hole to be $M_{BH}=7.07^{+2.05}_{-2.10}\times10^5$ and $M_{BH}=7.03^{+1.15}_{-1.19}\times10^5$ M$_\odot$, disc accretion rates $\Dot{m}_d=0.07\pm0.02$ and $0.08\pm0.01$ $\Dot{m}_{Edd}$, sub-Keplerian halo accretion rates $\Dot{m}_h=0.10^{+0.02}_{-0.02}$ and $\Dot{m}_h=0.09^{+0.01}_{-0.01}$ $\Dot{m}_{Edd}$ for both the observations, respectively. We find the shock locations at $18.38^{+5.60}_{-5.35}$ $r_g$ and $11.03\pm4.65$ $r_g$ with shock strengths of $1.77^{+0.56}_{-0.50}$ and $1.68^{+0.16}_{-0.19}$ for XRT3 and XRT4 observations, respectively. The parameters obtained from the {\tt TCAF} model fitting are presented in Table~\ref{tab:spec_fit_TCAF} and the model fitted spectra with the variation of $\chi$ are presented in the bottom middle and bottom right panels of Figure~\ref{fig:all_spec}.  

Through the comprehensive spectral fitting of the long-term ($2006-2021$) X-ray observations of UGC~6728 using {\tt TCAF model}, we encounter a negligible amount of extra-galactic hydrogen column density $(N_H)$. The estimated mass of the central black hole is $7.13\pm1.23~~\times~~10^5$ M$\odot$, which closely aligns with the mass estimated from the reverberation mapping method conducted by \citet{Bentz2016}. Furthermore, this model provides valuable insights into the dynamics of accretion flow surrounding the central black hole. 

We observe variations in the disc accretion rate $(\Dot{m}_d)$, ranging from $0.05\pm0.01$ $\Dot{m}_{Edd}$ to $0.08\pm0.01$ $\Dot{m}_{Edd}$, which has an increasing trend over time, reaching its maximum during the 2021 observation. The shock location, representing the boundary of the Compton cloud, moves accordingly. The Compton cloud is found to be the largest ($31.58^{+4.65}_{-3.56}$ ~$r_g$) during the 2006 XMM-Newton observation. Subsequently, the location of the shock gradually shifts inward. The halo accretion rate $(\Dot{m}_h)$ displays variations from $0.09\pm0.01$ $\Dot{m}_{Edd}$ to $0.11\pm0.01$ $\Dot{m}_{Edd}$, peaking during the 2017 (XRT2+N2) observation. Consequently, this observation exhibits a high X-ray luminosity for that time of observation. We also detect a narrow Fe~K$\alpha$ line at $\sim6.41$ keV with an equivalent width (EW) of approximately 100 eV in the XRT1+N1 (2016) and XRT2+N2 (2017) observations.

The variations of accretion flow parameters of {\tt TCAF} model, namely $\Dot{m}_d$, $\Dot{m}_h$, $X_s$, and $R$, are presented in Table~\ref{tab:spec_fit_TCAF} and potted in Figure~\ref{fig:parameter}, with other parameters. The unabsorbed X-ray luminosity in the $0.5-10.0$ keV range, determined using the {\tt `clumin'} task in {\tt Xspec} for all the observations, is provided in Table~\ref{tab:spec_fit_TCAF}. It is observed that the source remained in the sub-Eddington luminosity regime throughout the observation period. The long-term variations of the X-ray spectral parameters are also illustrated in Figure~\ref{fig:parameter}.

\begin{table*}
	  \caption{\label{tab:spec_fit_TCAF} Parameters obtained from the {\tt TCAF} model fitting with all the spectra. The unabsorbed X-ray luminosity is calculated for the energy range from $0.5$ to $10.0$ keV.}	
  	\vskip -0.2 cm
  	{\centerline{}}
  	\begin{center}
  		\begin{tabular}{c c c c c c c c c c c}
  			\hline
\rm ID & \rm MJD  &  \rm $N_H$          & \rm $M_{BH}$ & \rm $\Dot{m}_d$ & \rm $\Dot{m}_h$ & \rm $X_s$  & \rm $R$  & \rm $N$     &  \rm $\chi^2/dof$ & $\log(L_x)$  \\
       &          &  $(10^{20})$        &$(10^5 M_\odot)$ & $(\Dot{m}_{Edd})$& $(\Dot{m}_{Edd})$&  $(r_g)$  & &$(10^{-5})$ &  &  \\
\hline
&&&&&&& \\
 XMM  & 53789 & $0.12^{+0.10}_{-0.10}$ &$7.28^{+1.09}_{-1.08}$ & $0.05^{+0.01}_{-0.01}$ & $0.07^{+0.01}_{-0.01}$ & $31.58^{+4.65}_{-3.56}$ &  $1.92^{+0.99}_{-0.91}$ & $0.92^{+0.63}_{-0.65}$ & $214.12/209$& $42.14\pm0.25$\\
 &&&&& \\
 SU  & 54988 & $0.15^{+0.11}_{-0.12}$ &$7.09^{+0.59}_{-0.57}$ & $0.05^{+0.01}_{-0.01}$ & $0.08^{+0.01}_{-0.01}$ & $29.09^{+2.10}_{-2.09}$ &  $1.40^{+0.07}_{-0.09}$ & $7.40^{+0.95}_{-0.92}$ & $277.31/320$& $43.06\pm0.43$\\
 &&&&& \\
XRT1+N1 & 57579 & $0.10^{+0.10}_{-0.10}$ &$7.11^{+0.39}_{-0.32}$ & $0.06^{+0.01}_{-0.01}$ & $0.09^{+0.01}_{-0.01}$ & $23.62^{+2.55}_{-2.56}$ &  $1.68^{+0.09}_{-0.08}$ & $1.51^{+0.37}_{-0.38}$ & $252.10/232$& $41.78\pm0.22$\\
 &&&&& \\
XRT2+N2 & 58093 & $0.10^{+0.10}_{-0.10}$ &$7.14^{+0.22}_{-0.19}$ & $0.07^{+0.01}_{-0.01}$ & $0.11^{+0.01}_{-0.01}$ & $21.13^{+2.07}_{-1.56}$ &  $1.34^{+0.06}_{-0.05}$ & $1.81^{+0.80}_{-0.78}$ & $671.74/626$ & $42.16\pm0.17$\\
 &&&&& \\
XRT3 & 59097 & $0.09^{+0.10}_{-0.09}$ &$7.07^{+2.05}_{-2.10}$ & $0.07^{+0.02}_{-0.02}$ & $0.10^{+0.02}_{-0.02}$ & $18.38^{+5.60}_{-5.35}$ &  $1.77^{+0.56}_{-0.50}$ & $0.63^{+0.80}_{-0.88}$ & $22.40/22$ & $42.15\pm0.22$\\
 &&&&& \\
XRT4 & 59414 & $0.12^{+0.10}_{-0.12}$ &$7.03^{+1.15}_{-1.19}$ & $0.08^{+0.01}_{-0.01}$ & $0.09^{+0.01}_{-0.01}$ & $11.03^{+4.65}_{-4.65}$ &  $1.68^{+0.16}_{-0.19}$ & $0.98^{+0.80}_{-0.95}$ & $52.37/59$ & $42.14\pm0.15$\\
 &&&&& \\
\hline
  \end{tabular}
  	\end{center}
  \end{table*}

\subsubsection{relxillCp}
\label{relxill}

From the power-law fitting above 10.0 keV, we observed that the high-energy spectra of UGC~6728 lacked a reflection component in the high-energy domain above 10 keV. However, the presence of Fe-line (see Table~\ref{tab:cutpoffpl}) suggests the existence of a reflection component over the primary continuum above 10.0 keV. Despite the considerable uncertainty in the temperature estimation, we also observed a correlation between the equivalent width of the Fe-line and the temperature of the Compton cloud. It is anticipated that the equivalent width decreases as the temperature of the Compton cloud decreases unless the photons from outflows excite the Fe atoms \citep{CT1995}. To investigate the presence of reflection in the X-ray spectra of UGC~6728, we employ the relativistic reflection model RELXILL \citep{Garcia2013, Garcia2014, Dauser2013, Dauser2014, Dauser2016} on the broadband spectra. This model assumes a broken power-law emission profile with the broken radius $R_{br}$. The emissivity is expressed as $E(r)\sim r^{-q_1}$ for $r < R_{br}$ and $E(r)\sim r^{-q_2}$ for $r > R_{br}$, where $q_1$ and $q_2$ are the inner and outer emissivity index, respectively. The model parameter $R_{refl}$ provides information about the ratio of intensity emitted towards the disc compared to escaping to infinity \citep{Dauser2016}. Other parameters, such as the ionization parameter $(\xi)$, iron abundance $(A_{Fe})$, inclination angle $(i^0)$, inner radius of the accretion disc $(R_{in})$, and the spin of the black hole $(a)$, are allowed to vary during the fitting process.

For the broadband X-ray spectral fitting, we use {\tt relxillCp} flavour with the final model specified in {\tt Xspec} as {\tt Tbabs*zTbabs*constant*relxillCp}. The {\tt relxillCp} flavour does not assume any particular geometry, and the primary continuum emission is given by the thermal Comptonized model depending on the spectral index $(\Gamma)$ and cut-off energy $(E_c)$. We do not use additional Gaussian components to fit the Fe-line as this model incorporates Fe abundance as a free parameter. We fix the values of the outer radius of the accretion disc at $R_{out}=1000~r_g$ and redshift $z=0.0065$ and the corresponding results are represented in Table~\ref{tab:relxill}.

\begin{figure*}
		\centering
  \includegraphics[width=0.29\textwidth, angle =0]{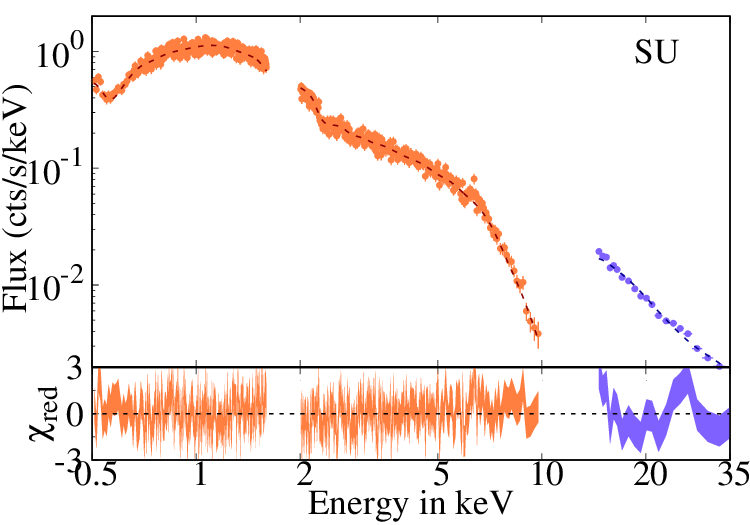}
  \includegraphics[width=0.29\textwidth, angle =0]{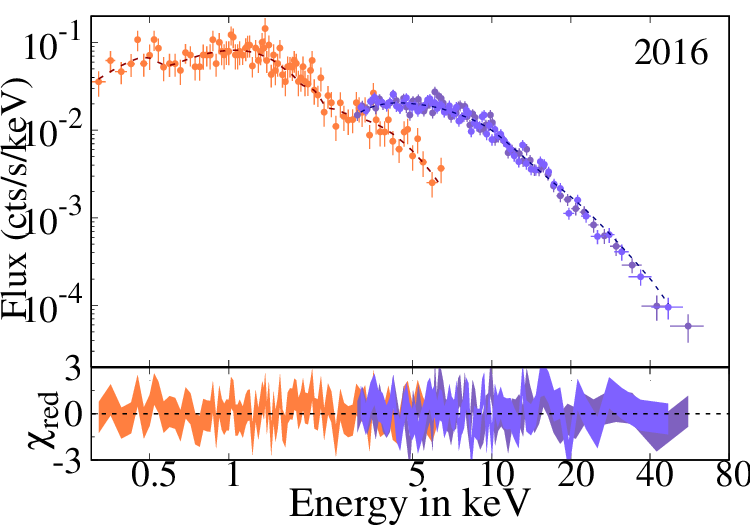}
  \includegraphics[width=0.29\textwidth, angle =0]{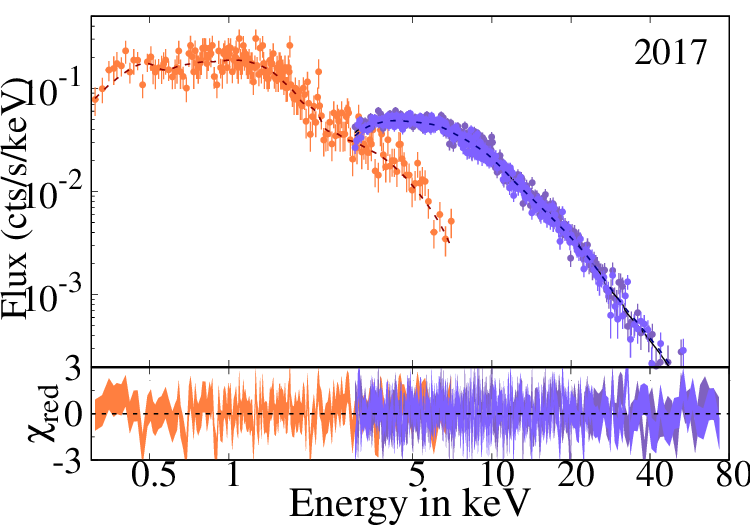}
  \caption{{\tt relxillCp} model fitted X-ray spectra of UGC~6728 from various observations from 2009--2021, along with the residuals obtained from the spectral fitting.}
		\label{fig:all_spec_relxill}
	\end{figure*}

We apply the {\tt relxillCp} model to SU observation (MJD--54988), and this model provided a satisfactory fit of the X-ray spectrum with $\chi^2/dof=621.96/563$. The parameters for the primary continuum are estimated as $\Gamma=1.50^{+0.04}_{-0.05}$ and $E_c=358.52^{+6.89}_{-19.5}$ keV. The inner and outer emissivity indices are found to be $3.00^{+0.84}_{-0.97}$ and $5.23^{+0.59}_{-0.52}$ for $R_{in}=3.94^{+1.60}_{-0.98}~r_g$ and $R_{br}=15.49^{+8.64}_{-8.97}~r_g$, respectively. The Fe abundance for this observation was estimated as $1.05^{+0.70}_{-0.74}~A_\odot$ with ionisation parameter $\log(\xi)=3.71^{+0.32}_{-0.33}$ and reflection fraction $R_{refl}=0.74^{+0.11}_{-0.10}$. The intrinsic parameters of this source were estimated as the black hole spin $a=0.99^{+0.12}_{-0.34}$ and inclination angle $i=48.20^{+14.2}_{-15.4}$ degrees. The spectrum is plotted on the left panel of Figure~\ref{fig:all_spec_relxill}. 

We followed a similar procedure for XRT1+N1 (MJD--57579) and XRT2+N2 (MJD--58093) observations for the broadband, and the corresponding estimated values from the X-ray spectral fitting is given in Table~\ref{tab:relxill}. For these observations, the parameters for the primary continuum were obtained as follows: the spectral index $\Gamma$ are $1.65^{+0.04}_{-0.03}$ and $1.70^{+0.02}_{-0.03}$ with cut-off energy $E_c$ are $273.68^{+29.56}_{-67.38}$ and $33.24^{+15.62}_{-56.24}$ keV, respectively. The accretion disc extended up to $R_{in}=6.14^{+1.21}_{-1.11}~r_g$ and $R_{in}=21.49^{+1.32}_{-1.33}~r_g$ with break radius $R_{br}=39.61^{+10.22}_{-10.18}~r_g$ and $R_{br}=15.0~r_g$, respectively. We fixed $R_{br}$ at its default value ($15.0~r_g$) for XRT2+N2 observation as this parameter was not well constrained during fitting. The emissivity indices are found at $q_1=2.85^{+0.47}_{-0.61}$ and $q_1=3.00^{+0.63}_{-0.58}$ respectively for inner region where $r < R_{br}$ and $q_2=3.01^{+0.52}_{-0.51}$ and $q_2=3.00^{+0.50}_{-0.59}$ respectively for outer region of these observations. The Fe abundance and the ionization parameters are nearly same and the values are  $A_{Fe}=0.51^{+0.20}_{-0.21}~A_\odot$ and $0.50^{+0.20}_{-0.23}~A_\odot$ with $\log(\xi)=3.45^{+0.63}_{-0.62}$ and $3.37^{+0.22}_{-0.23}$, respectively. The reflection fraction is determined to be $R_{refl}=0.11^{+0.05}_{-0.07}$ and $0.12^{+0.08}_{-0.05}$ for these observations. It is expected that the intrinsic parameters would be constant for all observations and the estimated values for black hole spin and inclination angles are $a=0.98^{+0.20}_{-0.18}$ and $a=0.96^{+0.29}_{-0.30}$ with $i=53.59^{+15.7}_{-16.1}$ and $i=46.62^{+13.6}_{-12.3}$ degrees, respectively. The spectra are plotted on the middle and right panels of Figure~\ref{fig:all_spec_relxill}.

\begin{table}
	  \caption{\label{tab:relxill} Parameters obtained from the {\tt relxillCp} model fitting with all the spectra. }	
  	\vskip -0.2 cm
  	{\centerline{}}
  	\begin{center}
  		\begin{tabular}{c c c c}
  			\hline
\rm Model Parameters        & \rm SU           &  \rm XRT1+NU1     & \rm XRT2+NU2   \\
                            & \rm (MJD--54988) & \rm  (MJD--57579) & \rm (MJD--58093) \\
    \hline
    \hline
    $N_H(10^{20}~cm^{-2})$  &$0.23^{+0.21}_{-0.20}$ &$0.75^{+0.20}_{-0.22}$ &$0.73^{+0.25}_{-0.23}$ \\
    &&&\\
    $q_1$                    &$3.00^{+0.84}_{-0.97}$ &$2.85^{+0.47}_{-0.61}$ &$3.00^{+0.63}_{-0.58}$ \\
    &&&\\
    $q_2$                    &$5.23^{+0.59}_{-0.52}$ &$3.01^{+0.52}_{-0.51}$ &$3.00^{+0.50}_{-0.59}$ \\
    &&&\\
    $R_{br}(r_g)$           &$15.49^{+8.64}_{-8.97}$ &$39.61^{+10.22}_{-10.18}$ &$15.00^{f}$ \\
    &&&\\
    $R_{in}(r_g)$           &$3.64^{+1.60}_{-0.98}$ &$6.14^{+1.21}_{-1.11}$ &$5.66^{+1.32}_{-1.33}$ \\
    &&&\\
    $a$                     &$0.99^{+0.12}_{-0.34}$ &$0.98^{+0.20}_{-0.18}$ &$0.96^{+0.29}_{-0.30}$ \\
    &&&\\
    $i(^0)$                     &$48.20^{+14.2}_{-15.4}$ &$53.59^{+15.7}_{-16.1}$ &$46.62^{+13.6}_{-12.3}$ \\
    &&&\\
    $\Gamma$                    &$1.50^{+0.04}_{-0.05}$ &$1.65^{+0.04}_{-0.03}$ &$1.70^{+0.02}_{-0.03}$ \\
    &&&\\
    $\log \xi$                  &$3.71^{+0.32}_{-0.33}$ &$3.45^{+0.63}_{-0.62}$ &$3.37^{+0.22}_{-0.23}$ \\
    &&&\\
    $A_{fe}(A_\odot)$           &$1.05^{+0.70}_{-0.74}$ &$0.51^{+0.20}_{-0.21}$ &$0.50^{+0.20}_{-0.23}$ \\
    &&&\\
    $kT_e(keV)$                 &$358.52^{+6.89}_{-19.5}$ &$273.68^{+29.56}_{-67.38}$ &$233.24^{+15.62}_{-56.24}$ \\
    &&&\\
    $R_{refl}$                 &$0.74^{+0.11}_{-0.10}$ &$0.11^{+0.05}_{-0.07}$ &$0.12^{+0.08}_{-0.05}$ \\
    &&&\\
    $N(10^{-5})$           &$12.2^{+1.6}_{-1.1}$ &$3.94^{+0.58}_{-0.58}$ &$9.01^{+0.35}_{-0.30}$ \\
\hline
$\chi^2/dof$               & 621.96/563      & 223.48/224  &800.86/802\\

\hline
  \end{tabular}
  $^f$ indicates the parameter that was fixed during X-ray spectrum fitting. 

  	\end{center}
  \end{table}


\subsection{Timing Analysis}
\label{sec:timing}
\begin{figure*} 
	\centering
	\includegraphics[trim={0 2.5cm 0 0}, scale=1.3]{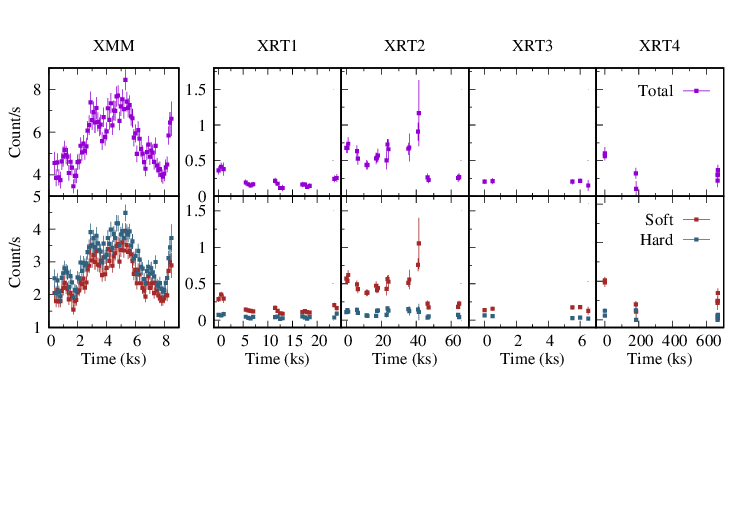}
        \caption{Variation of photon counts with respect to time from {\it XMM-Newton} and {\it Swift}/XRT observations of UGC~6728 at different epochs (see Table~\ref{tab:1}). The light curves in the $0.3-$10.0 keV range are shown in the top panels, whereas the bottom panels show the light curves in the soft X-ray band (represented by red dots/line) and hard X-ray band (represented by blue dots/line), plotted together for comparison. The top and bottom left panels represent light curves from the {\it XMM-Newton} observation, whereas the other panels are from the {\it Swift}/XRT observations of UGC~6728. }
	\label{fig:lc} 
\end{figure*}

We conducted a comprehensive timing analysis of the light curves with time binsize=$500$s obtained from the {\it XMM-Newton}, {\it Swift}, and {\it NuSTAR} observations of UGC~6728 (see Table~\ref{tab:1}). To investigate the variability, we partitioned the X-ray light curves from the {\it XMM-Newton} and {\it Swift} observations in the 0.3-10 keV energy range into two segments: the soft excess component (0.3-2.0 keV range) and the primary continuum component (3.0-10.0 keV range). The light curves in the 0.3-10 keV range are illustrated in the top panels of Figure~\ref{fig:lc}. The bottom panels in the figure show the light curves of the source in the soft and hard X-ray ranges. The left panels represent the light curves from the {\it XMM-Newton} observation, whereas the other panels represent the light curves from the {\it Swift}/XRT observations of UGC~6728. In this figure, it becomes apparent that the source count rates during the {\it XMM-Newton} observation significantly exceed the count rates observed during the XRT observations. This discrepancy may be attributed to the use of different instruments for the respective observations.

To explore the variability in the high-energy regime, we considered light curves from {\it NuSTAR} observations. For the correlation study in the broad energy range, we considered the {\it NuSTAR} observations in 2016 (N1) and 2017 (N2) due to their substantial exposure times (above 20 ks; see Table~\ref{tab:1}). The light curves obtained from the {\it NuSTAR} observations in the $3.0-79.0$ keV energy range were divided into two energy bands: $3.0-6.0$ keV and $7.0-79.0$ keV ranges and were utilized for the correlation study (see Figure~\ref{fig:delay}). Notably, we excluded the light curve in the $6.0-7.0$ keV band as it predominantly comprises photons associated with the Fe K$_\alpha$ emission line, while our primary interest lies in investigating the continuum emission.

\subsubsection{Variability}
\label{sec:fvar}
X-ray variability is a powerful tool to probe the nearby regions of the central supermassive black bole of an AGN. Since UGC~6728 has a `bare-nucleus' (see Table~\ref{tab:spec_fit_Nthcomp} and Table~\ref{tab:spec_fit_TCAF}), the X-ray photons originated at the Compton cloud are not intercepted by any other medium such as BLR, NLR or molecular torus. As a result, the variability in the X-ray emission of this source is very likely produced by the Compton cloud itself or the very inner edge of the accretion disc. To study the temporal variability of UGC~6728, we estimated different variability parameters for all the X-ray observations used in the present work. The results are presented in Table~\ref{tab:1}. The fractional variability $F_{var}$ \citep{Edelson1996, Nandra1997, Rodriguez1997, Edelson2001, Vaughan2003, Edelson2012} for a light curve of $x_i$ counts/s with error $\sigma_i$ for $N$ number of data points, with a mean count rate $\mu$ and standard deviation $\sigma$ is given by,
\begin{equation}
    F_{var}=\sqrt{\frac{\sigma^2_{XS}}{\mu^2}}
\end{equation}
where, $\sigma_{XS}$ is the excess variance \citep{Nandra1997, Edelson2002} which is defined as,
\begin{equation}
    \sigma^2_{XS}=\sigma^2 -\frac{1}{N}\sum_{i=1}^{N} \sigma^2_i.
\end{equation}

We defined the normalized excess variance as $\sigma^2_{NXS}=\sigma^2_{XS}/\mu^2$ and the uncertainties on corresponding parameters $(\sigma^2_{NXS}, \text{and} F_{var})$ are calculated as described by \citep{Vaughan2003, Edelson2012}. To investigate the variability in the X-ray light curves, we calculated peak-to-peak amplitude as $R=x_{max}/x_{min}$, where $x_{max}$ and $x_{min}$ are the maximum and minimum flux, respectively. 

The X-ray light curves from different energy bands exhibit different magnitudes of variability. The results from the variability analysis are presented in Table~\ref{tab:fvar}. During 2006 {\it XMM-Newton} observation, the highest count rates were observed in all energy bands ranging from $0.3$ to $10.0$ keV. For this entire energy range ($0.3$-$10.0$ keV), the maximum count rate $(x_{max})$, the minimum count rate $(x_{min})$ and the mean count rate $(\mu)$ were $8.44$ counts/s, $3.47$ counts/s, and $5.60$ counts/s, respectively. For this observation, the soft band (0.3-3.0 keV range) mean count rate $(\mu)$ was determined to be $2.58$ counts/s, with a maximum count rate $(x_{max})$ and minimum count rate $(x_{min})$ of $3.95$ counts/s and $1.55$ counts/s, respectively. Along with this, the primary continuum (3.0-10.0 keV) exhibited a similar trend with $\mu$, $x_{max}$ and $x_{min}$ of $3.02$ counts/s, $4.50$ counts/s, and $1.92$ counts/s, respectively. The variability parameters such as the peak-to-peak amplitude $(R)$, normalized excess variance $\sigma^2_{NXS}$, and fractional variability $(F_{var})$ are found to be comparable for the light curves in the low ($0.3-2.0$ keV), high ($3.0-10.0$ keV) and total ($0.3-10.0$ keV) energy ranges of this observation. We found that $R$ varies in the range of $2.32$ to $2.56$, whereas $\sigma^2_{NXS}$ and $F_{var}$ are approximately the same at $0.04\pm0.004$ and $\sim20\%$, respectively, for the low-energy, high energy, and the entire energy bands. The detailed results are given in Table~\ref{tab:fvar}.


In the case of {\it Swift}/XRT observations, the exposure time is comparatively lower with respect to other observations (see Table~\ref{tab:1}). As a result, we have a smaller number of data points for these observations (see Table~\ref{tab:fvar} and Figure~\ref{fig:lc}). In the soft energy band ($0.3-2.0$ keV), we found that the maximum count rates ranged from $0.35$ count/s to $1.05$ count/s, with corresponding minimum count rates varying from $0.09$ count/s to $0.17$ count/s. The mean count rates ranged from $0.16$ count/s to $0.48$ count/s. We observe a similar trend in the high energy band ($3.0-10.0$ keV), with maximum count rates ranging from $0.40$ count/s to $0.61$ count/s, minimum count rates varying from $0.01$ count/s to $0.18$ count/s, and mean count rates ranging from $0.19$ count/s to $0.30$ count/s. For the entire energy range ($0.3-10.0$ keV) of the {\it Swift}/XRT observations, the maximum count rates varied from $0.22$ count/s to $1.16$ count/s, minimum count rates ranged from $0.1$ count/s to $0.23$ count/s, and the mean count rates varied from $0.20$ count/s to $0.58$ count/s. In all the {\it Swift}/XRT observations, we observed that XRT2 had comparatively higher counts compared to other observations, as shown in Figure~\ref{fig:lc} and presented in Table~\ref{tab:fvar}. Due to the low count rate and high error associated with each data point,  we encountered negative values for normalized excess variance ($\sigma^2_{NXS}$), resulting in imaginary fractional variability ($F_{var}$) for most of the XRT observations. However, for observations with positive $\sigma^2_{NXS}$, we observed fractional variability ($F_{var}$) ranging from $35.4\%$ to $43.5\%$, with an average of $39.5\%$ for the low energy band, and from $30.1\%$ to $38.8\%$, with a mean of $35.0\%$ for the entire energy band. In the energy range from 3.0 to 10.0 keV, only XRT3 showed a positive $\sigma^2_{NXS}=(0.35\pm0.07)$ with $F_{var}=18.7\pm2.1\%$. The detailed results are presented in Table~\ref{tab:fvar}. The soft X-ray variability studies could be improved with future missions, such as \textit{Colibr\'i} \citep{Heyl2019, Caiazzo2019, Heyl2020}.

We have only two {\it NuSTAR} observations of UGC~6728 in 2016 (N1) and 2017 (N2). The entire energy band ($3.0-79.0$ keV) is divided into several sub-energy bands, namely, $3.0-6.0$ keV for the primary continuum, $6.0-7.0$ keV for the Fe K$_\alpha$ line and $7.0-79.0$ keV for the high energy counter-part. Figure~\ref{fig:delay} displays the temporal variation of the X-ray photons in the energy bands of $3.0-6.0$ keV and $7.0-79.0$ keV for the 2016 (N1) and 2017 (N2) observations. The results of the variability analysis for different energy bands of these observations can be found in Table~\ref{tab:fvar}. For the N1 observation, we observed a maximum count rate of $0.20$ count/s and a minimum count rate of $0.04$ count/s in the $3.0-6.0$ keV energy band. The average count rate ($\mu$) in this band was $0.10$ count/s. A similar count rate pattern was observed in the high-energy counterpart ($7.0-79.0$ keV), with maximum, minimum and average count rates of $0.35$ count/s, $0.09$ count/s, and $0.18$ count/s, respectively. The Fe K$\alpha$ energy band ($6.0-7.0$ keV) exhibited lower count rates, with maximum count rates of $0.08$ count/s, minimum count rates of $0.01$ count/s, and an average count rate of $0.04$ count/s. For the entire X-ray energy band ($3.0-79.0$ keV), the count rates ranged from a maximum value of $0.53$ count/s to a minimum value of $0.17$ count/s, with a mean value of $0.32$ count/s. The variability parameter ($F{var}$) ranged from $11.6\pm5.2\%$ to $31.6\pm2.6\%$, indicating higher variability in the lower energy band compared to the higher energy band for this observation. In the entire energy range, the average variability was $F_{var}=22.9\pm5.6\%$. Please refer to Table~\ref{tab:fvar} for detailed results.

In the case of the 2017 {\it NuSTAR} observation (N2), we observed a similar trend to N1. For the lower energy band ($3.0-6.0$ keV), the maximum, minimum and average count rates were $0.37$ count/s, $0.07$ count/s, and 0.22 count/s, respectively. In the Fe K$\alpha$ energy band ($6.0-7.0$ keV), the photon counts varied from a maximum value of $0.12$ count/s to a minimum value of $0.02$ count/s, with an average count rate of $\mu=0.07$ count/s. However, for the high-energy counterpart ($7.0-79.0$ keV), the errors associated with each photon count were higher than the standard deviation of the light curve. Consequently, the fractional variability $F{var}$ became imaginary, and those values are not presented in Table~\ref{tab:fvar}. For this light curve, we found that the $x_{max}=0.05$ count/s and $x_{min}=0.02$ count/s with $\mu=0.03$ count/s. For the entire energy range ($3.0-79.0$ keV), the maximum count rate was $1.17$ count/s, the minimum count rate was $0.24$ count/s, and the average count rate was $\mu=0.62$ count/s. We found that the variability parameter $F_{var}$ remained nearly constant for all energy bands as well as for the entire energy range, ranging from $21.8\pm4.1\%$ to $26.5\pm2.0\%$. The detailed results are presented in Table~\ref{tab:fvar}.

\begin{table*}
\centering
\caption{The table presents the variability statistics in different energy ranges for various observations for 500s time bins. It is important to note that in some cases, the average error of the observational data exceeds the $1\sigma$ limit, leading to negative excess variance. Consequently, these cases have imaginary values for $F_{var}$, which are not included in the table.}
\begin{tabular}{lcccccccc}
\hline
ID      & Energy band     &    $N$   &$x_{max}$ &$x_{min}$& $\mu$ &$R=\frac{x_{max}}{x_{min}}$   &$\sigma^2_{NXS}$      &$F_{var}   $\\
&&&&&&\\
	&   keV           &          & Count/s  & Count/s & Count/s &    &           & $(\%)$\\
\hline
XMM     &0.3-2.0          &    82     & 3.95     & 1.55 & 2.58  &2.56 & $0.04\pm0.004$       & $21.2\pm1.25$ \\
XMM     &3-10.0           &    82     &4.50      &1.92  & 3.02   &2.32 &$0.04\pm0.003 $       & $19.1\pm2.11$ \\
XMM     &0.3-10.0         &    82    &8.44      &3.47   & 5.60   &2.44  & $0.04\pm0.003$       & $20.1\pm1.10$ \\
\hline
XRT1    &0.3-2.0          &    17     & 0.35     & 0.09 & 0.16   &3.92 & $0.18\pm0.036$       & $43.5\pm2.35$ \\
XRT1    &3-10.0           &    17     &0.54      &0.18  & 0.30   &3.03 &$-0.03$               & $--$ \\
XRT1    &0.3-10.0         &    17    &0.41      &0.12   & 0.21   &3.60  & $0.14\pm0.038$        & $38.0\pm2.01$ \\
\hline
XRT2    &0.3-2.0          &    21     & 1.05     & 0.17 & 0.48   &6.01 & $0.13\pm0.032$       & $35.4\pm2.14$ \\
XRT2    &3-10.0           &    21     &0.40      &0.11  & 0.23   &9.71 & $-0.021$             & $--$ \\
XRT2    &0.3-10.0         &    21    &1.16      &0.23   & 0.58   &5.10  & $0.09\pm0.03$       & $30.1\pm2.00$ \\
\hline
XRT3    &0.3-2.0          &    05     & 0.18     & 0.13 & 0.16   &1.40 &$-0.24$               & $--$ \\
XRT3    &3-10.0           &    05     &0.48      &0.17  & 0.27   &2.92 & $0.35\pm0.07$        & $18.7\pm2.1$ \\
XRT3    &0.3-10.0         &    05    &0.22      &0.15   & 0.20   &1.43  & $-0.39$             & $--$ \\
\hline
XRT4    &0.3-2.0          &    08     & 0.51     & 0.10 & 0.29   &4.91 & $0.16\pm0.061$       & $39.5\pm2.20$ \\
XRT4    &3-10.0           &    08     &0.61      &0.01  & 0.19   &60.1 & $-0.05$             & $--$ \\
XRT4    &0.3-10.0         &    08    &0.61      &0.10   & 0.35   &5.78  & $0.150\pm0.061$      & $38.8\pm2.3$ \\
\hline
N1      &3.0-6.0          &    51    &0.20     &0.04    & 0.10  &5.45  & $0.101\pm0.019$      & $31.6\pm2.6$ \\
N1      &6.0-7.0          &    51    &0.08     &0.01    & 0.04  &7.30  & $0.019\pm0.015$      & $13.7\pm2.4$ \\
N1      &7.0-79.0        &    51    &0.35     &0.09    & 0.18  &3.80  & $0.014\pm0.010$      & $11.6\pm5.2$ \\
N1      &3.0-79.0         &    51    &0.53     &0.17    & 0.32  &3.13  & $0.052\pm0.012$      & $22.9\pm5.6$ \\

\hline
N2      &3.0-6.0          &    131    &0.37     &0.07   & 0.22   &5.61  & $0.070\pm0.006$      & $26.5\pm2.0$ \\
N2      &6.0-7.0          &    131    &0.12     &0.02   & 0.07   &4.77  & $0.048\pm0.009$      & $21.8\pm4.1$ \\
N2      &7.0-79.0         &    129    &0.05     &0.02   & 0.03   &2.10  & $-0.98$      & $--$ \\
N2      &3.0-79.0         &    131    &1.17     &0.24   & 0.62   &4.91  & $0.062\pm0.003$      & $25.0\pm02.0$ \\
\hline
\end{tabular}
\label{tab:fvar}
\end{table*}

\subsubsection{Correlation}
\label{sec:zdcf}
For the temporal analysis of the long-term X-ray observation of UGC~6728, we concentrated on two epochs of {\it NuSTAR} observations: 2016 and 2017 which were accompanied by low energy ($0.3-10.0$ keV) observations from {\it Swift}/XRT. To study the correlation between the light curves, we have chosen the simultaneous observations from {\it NuSTAR} (N1 and N2) in two energy bands: (i) $3.0-6.0$ keV and (ii) $7.0-79.0$ keV. We excluded the light curve in $6.0-7.0$ keV band as it contains primarily photons from the Fe K$_\alpha$ line, which appears at $\sim6.4$ keV (see Section~\ref{sec:spectral}). 

\begin{figure} 
	\centering
	\includegraphics[scale=0.7]{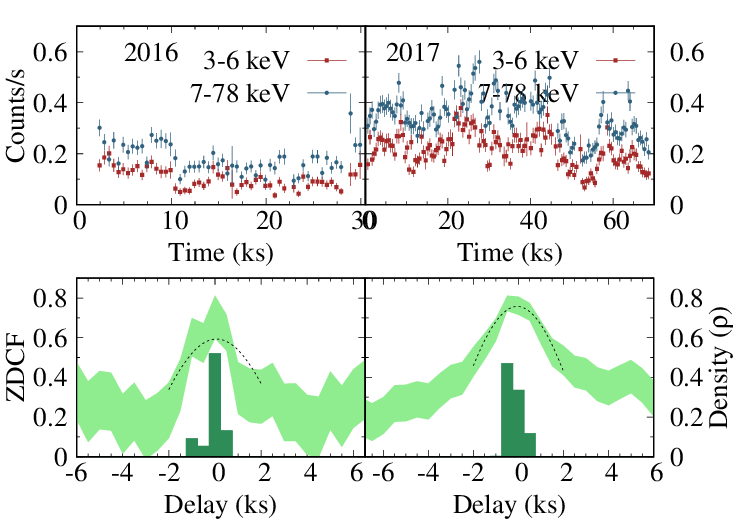}
	\caption{Top panel: The light curves of UGC~6728 in the energy bands of $3.0$ to $6.0$ keV (red) and $7.0$ to $79.0$ keV (blue) are presented for two epochs of {\it NuSTAR} observations in 2016 and 2017. The high energy band have higher count rates in comparison to the low energy band for these two observations. Lower Panel: $\zeta$-discrete cross-correlation functions (light-green) are plotted for the light curves in $3.0-6.0$ keV and $7.0-79.0$ keV ranges. The likelihood functions (dark-green), simulated using $15000$ points, are plotted along with the {\tt ZDCF}.}
	\label{fig:delay} 
\end{figure}

The $\zeta-$discreate cross-correlation function ({\tt ZDCF}\footnote{\url{www.weizmann.ac.il/particle/tal/research-activities/software}}, \citep{Alexander1997}) was utilized for this analysis. The likelihood of the correlation function was calculated for the same {\tt DCF} function using $15000$ simulated points in the {\tt ZDCF} code for the light curves obtained from the {\it NuSTAR} observations. We fitted a {\tt Gaussian} function to estimate the peak and the error on the peak is obtained using the formula provided by \citet{Gaskell1987}. Our approach followed a similar procedure as described in \citet{Nandi2021}. Initially, we fitted all the light curves with a straight line and found the $\chi^2_{red}<1.5$ with different slopes and intercepts. However, these data points are not suitable for linear fitting. Next, we performed the correlation analysis of these light curves to ensure the simultaneity of their observations (see the upper panel of Figure~\ref{fig:delay}). 

The {\tt ZDCF} obtained from both {\it NuSTAR} observations (N1 and N2) show a similar type of delay pattern. In the case of 2016 {\it NuSTAR} observation (N1), a delay of $66.8\pm500$s is observed. We fitted a {\tt Gaussian} function (indicated by the dotted black line in the lower panel of Figure~\ref{fig:delay}). The uncertainty of the position of the peak is obtained as $\Delta\tau_d=162.5$s, which is smaller than the time bin size (500s) of the light curves. Hence, we considered $500$s as the error of the position of the peak. The detailed results are given in Table~\ref{tab:delay}. We calculated the likelihood function, represented by deep-green bars in the lower panel of Figure~\ref{fig:delay}, to accurately determine the position of the peak. A similar procedure was followed for the 2017 observation (N2). In this case, we found that the delay between the two given energy bands is $\tau_d^{zdcf}=66.8$s. The uncertainty of the position of the peak, obtained from the prescription provided by \citep{Gaskell1987} is $\Delta\tau_d=46.5$s, is much smaller than the time bin size. As with the N1 observation, we considered the time bin size ($500$s) as the uncertainty for the parameter $\tau_d^{zdcf}$. The results are summarized in Table~\ref{tab:delay}.

\begin{table}
\centering

    \caption{The parameters used for estimating the time delay are presented. The peak value of the correlation function is denoted by $\epsilon^z_\tau$, and the corresponding time delay is represented by $\tau^{zdcf}_{d}$. The uncertainty of the position of the peak of the correlation function is calculated using the procedure described in \citet{Gaskell1987} and is given by $\Delta\tau_{d}$. To ensure the reliability of the results, we compared this error with the time bin size and selected the larger one for our analysis. Further details can be found in  Section~\ref{sec:zdcf}.}    
    \label{tab:delay}

    \begin{small}
    \begin{tabular}{|c|c|c|c|c|c}
    \hline
Id & Epochs& Bin size& $\Delta\tau_{d}$ &$\epsilon^{z}_{\tau}$  &{\bfseries $\tau^{zdcf}_{d}$}\\
  &{Year}      &  {(s)}  & (s)    &   &{(s)}       \\
    \hline                                                
    {N1}  &  2016 & 500 & 162.5 & $0.72\pm0.10$ & $ 66.8\pm 500$  \\
    {N2}  &  2017 & 500 & 46.5  & $0.78\pm0.06$   &$-50.9 \pm 500$         \\
     \hline
    \end{tabular}
    \end{small}
\end{table}

\section{discussions}
\label{sec:discussions}
In this study, we investigated the X-ray observations of UGC~6728 utilizing data from {\it XMM-Newton}, {\it NuSTAR} and {\it Swift}/XRT spanning from 2006 (MJD-53789) to 2021 (MJD-59414) (see Table~\ref{tab:1}).
We observed that UGC~6728 is a highly spinning $(a=0.98\pm0.3)$ black hole with an inclination angle around $\sim50$ degrees. The results of our spectral and temporal analysis are presented in Section~\ref{sec:result}. In this section, we will discuss the key findings derived from our data analysis.

\subsection{Evolution of Primary Continuum}
The `bare-nucleus' \citep{Walton2013} of UGC~6728 was observed over the period of 2006 to 2021 by various X-ray observatories, revealing both spectral and temporal variabilities. Throughout the long-term X-ray observations, the X-ray luminosity of UGC~6728 is found to vary in the range of $\sim10^{41.8}-10^{44.2}$ erg/s. The spectral index of the primary continuum ($3.0-10.0$ keV) also displayed variations ranging from $1.38\pm0.14$ to $1.85\pm0.05$, during this observational period. Correspondingly, the electron temperature of the Compton cloud varied from $475.89\pm250.35$ keV to $178.32\pm126.32$ keV. The hardest spectrum was observed in 2006 {\it XMM-Newton} observation, with a spectral index of $\Gamma=1.38\pm0.14$ and an electron temperature of $475.9\pm250.4$ keV according to the {\tt Nthcomp} model. This is the highest temperature of the electron cloud obtained from all the observations. It is worth noting that the spectral index $\Gamma$ is lower than that of other sources mentioned in the literature. This could be attributed to the relatively poor data quality, as the exposure time was limited for this observation. If we take into consideration the uncertainty of this parameter, its value falls within the range of $1.14$ to $1.52$, which is consistent with the observed $\Gamma$ values reported in the literature \citep{Kriss1980}. The corresponding temperature of the electron also appears to be high, with a large uncertainty for this observation. It is important to mention that the parameter, $kT_e$, is estimated using various models such as cut-off power-law, Nthcomp, and CompTT. The significant uncertainties in the estimation of $kT_e$ might be due to the absence of high-energy data beyond 10.0 keV. The model {\tt TCAF} provides insights into the accretion flow dynamics around the central object for this observation. By fitting the $0.2$ to $10.0$ keV spectra with the {\tt TCAF} model, we found that the sub-Keplerian halo accretion rate ($\Dot{m}_h=0.07\pm0.01$) was higher than the disc accretion rate ($\Dot{m}_d=0.05\pm0.01$) during that time. This suggests that the flow was sub-Keplerian-dominated for that observational period. This behavior is commonly observed in AGNs \citep{Nandi2019, Nandi2021}. We found that the boundary of the Compton cloud or the shock location $(X_s)$ was estimated to be  $\sim31.6$ $r_g$ with a relatively weak shock strength of $R\sim1.92$ based on spectral fitting. This is the highest shock strength when all the X-ray observations of this source are considered. A higher value of $R$ indicates the higher temperature of the Compton cloud, which is consistent with the results obtained from the {\tt Nthcomp} model fitting, implying an outward push of the boundary of the Compton cloud.

In the next {\it Suzaku} observation in 2009, we observe that the X-ray luminosity of UGC 6728 is maximum. This increase in luminosity can be attributed to the high normalization of the primary continuum during this observation. It is noted that the spectral index changed from $1.38\pm0.14$ to $1.50\pm0.07$, and the corresponding electron temperature decreased from $475.9\pm250.4$ to $350.42\pm151.74$ keV. This variation in spectral index and electron temperature could be explained by the variation in accretion dynamics. During this observation, we found that the accretion rates remained constant with those of the previous observation. However, the shock strength decreased from $1.92$ to $1.40$. As a result, the electrons in the Compton cloud could not gain energy from the shock, leading to a decrease in temperature. Moreover, we observed a prominent Fe-line at $6.54\pm0.1$ keV with an equivalent width of $188\pm17$ eV. The presence of the Fe-line suggests the existence of either reflection or a high abundance of iron in the accreting medium. To distinguish between these possibilities, we applied the {\tt relxillCp} model to this observation and found that the high abundance $(A_{Fe}=1.05\pm0.72)$ of iron with a relatively small amount of reflection $(R_{refl}=0.74\pm0.11)$.

We conducted a broadband ($0.3-79.0$ keV) spectral analysis of UGC~6728 using combined data from the 2016 {\it Swift}/XRT (XRT1) and {\it NuSTAR} (N1) observations. For this particular observation, we observed a relatively softer spectrum ($\Gamma=1.64\pm0.03$) for the primary continuum ($3.0-79.0$ keV) compared to the 2006 (XMM) observation.  The corresponding temperature of the electron cloud in the Compton cloud also changed from $475.9\pm250.35$ keV to $254.4\pm86.21$ keV. We determined the optical depth of the Compton cloud to be $0.69\pm0.02$, indicating an optically thin cloud. The disc accretion rate $(\Dot{m}_d)$ and the halo accretion rate $(\Dot{m}_h)$, changed from $0.05\pm0.01$ $\dot{M}_{Edd}$ to $0.06\pm0.01$ $\dot{M}_{Edd}$ and $0.07\pm0.01$ $\dot{M}_{Edd}$ to $0.09\pm0.01$ $\dot{M}_{Edd}$, respectively, with respect to the XMM observation. The increase in $\Dot{m}_d$ implies an increased supply of soft photons, which interact with the Compton cloud or corona, resulting in a relatively softer spectrum. Consequently, the shock location moved inward, with $X_s$ shifting from $31.58^{+4.65}_{-3.56}$ $r_g$ to $23.62\pm2.55$ $r_g$. The strength of the shock also increased from $1.40\pm0.08$ to $1.68\pm0.09$. These changes contributed to the observation of a relatively softer spectrum during this particular observation. By fitting a single {\tt Gaussian}, we detected a narrow Fe K$_\alpha$ line at $6.40\pm0.02$ keV with an equivalent width (EW) of $104\pm9$ eV. The detection of Fe-line suggests the presence of reprocessed emission. To investigate this, we applied the {\tt relxillCp} model on the broadband spectra. The model output indicated a reflection fraction of $R_{refl}=0.11\pm0.06$ along with an Fe abundance of $A_{Fe}=0.51\pm0.20$ $A_\odot$.

In the 2017 observation, which was carried out by {\it Swift}/XRT (XRT2) and {\it NuSTAR} (N2) simultaneously, we obtained another broadband spectrum $(0.3-79.0~\text{keV})$ with a higher X-ray luminosity compared to the previous broadband observation in 2016 (see the first panel of Figure~\ref{fig:lc}). By fitting the primary continuum ($3.0-10.0$ keV)  of this broadband spectrum with the {\tt Nthcomp} model, we found a marginal increase in the spectral index from $1.64$ to $1.69$, along with a decrease in the electron temperature $(kT_e)$ of the Compton cloud from $256.40\pm86.21$ keV to $249.18\pm41.95$ keV, with large uncertainties in $kT_e$. We also found the presence of an optically thin Compton cloud with optical depth $0.66\pm0.02$ during this observation. Additionally, we detected a narrow Fe line near 6.41 keV with an equivalent width (EW) of $80.9\pm8$ eV. To investigate the presence of reflection component beyond 10.0 keV, we used {\tt relxillCp} model to fit the broadband spectra and found that $A_{Fe}=0.50\pm0.2$ $A_\odot$ and $R_{refl}=0.11\pm0.6$. The low values of these parameters indicate the high energy spectrum is barely dominated by reflection above 10.0 keV.

The increase in X-ray luminosity suggests a change in the accretion flow, which can be quantified through {\tt TCAF} model fitting. We observed the accretion rates, $\Dot{m}_d$ and $\Dot{m}_h$, increase from $0.06\pm0.01$ $\dot{M}_{Edd}$ to $0.07\pm0.01$ $\dot{M}_{Edd}$ and $0.09\pm0.01$ $\dot{M}_{Edd}$ to $0.11\pm0.01$ $\dot{M}_{Edd}$, respectively. As a result, the luminosity of UGC~6728 increased by $\sim2.4$ times from the previous (2016) observation. Moreover, the shock location shifted inward from $23.62\pm2.56$ $r_g$ to $21.23\pm2.07$ $r_g$, accompanied by a corresponding decrease in the shock strength from $1.68\pm0.038$ to $1.34\pm0.80$.

In the 2020 observation carried out by {\it Swift}/XRT (XRT3), we encountered a spectrum similar to that of the previous observation in 2017. The flow parameters, as well as the spectral index ($\Gamma$) and electron temperature ($kT_e$), remained unchanged compared to the parameters obtained from the 2017 observation. Specifically, we found $\Gamma=1.66$ with $kT_e=249.18$ keV and corresponding optical depth $(\tau)$ is $0.69\pm0.02$. We also estimated the X-ray luminosity in $0.5-10.0$ keV range and found that this is nearly the same as in the case of the 2017 observation (see Table~\ref{tab:spec_fit_Nthcomp}). 

The accretion dynamics exhibited minimal changes in this observation. The {\tt TCAF} fitting suggests the accretion rates for 2020 observation to be $\Dot{m}_d=0.07\pm0.02$ $\dot{M}_{Edd}$ and $\Dot{m}_h=0.11\pm0.01$ $\dot{M}_{Edd}$ which are similar to those during the 2017 observation. The shock location is located near $18.38\pm5.6$ $r_g$ with a shock strength of $1.77\pm0.56$ for this observation. 

In the 2021 observation from {\it Swift}/XRT (XRT4), we found the softest X-ray spectrum for UGC~6728. The spectral index was determined to be $\Gamma=1.85\pm0.05$, with an electron temperature of $kT_e=178.32\pm126.32$ keV. The corresponding optical depth was measured to be $\tau=0.71\pm0.02$. During this observation, we estimated the maximum disc accretion rate as $\Dot{m}_d=0.08\pm0.01$ $\dot{M}_{Edd}$. Additionally, the halo accretion rate decreased from $0.109\pm0.01$ $\dot{M}_{Edd}$ to $0.09\pm0.01$ $\dot{M}_{Edd}$.

The increase in the disc accretion rate leads to the cooling of the Compton cloud due to the presence of soft photons generated in the disc. As a result, the Compton cloud cannot maintain its size and gradually shrinks. This causes the shock location to move inward, eventually settling at $11.03\pm4.65$ $r_g$, with a shock strength of $R=1.68\pm0.2$. The reduction in the size of the Compton cloud affects the production of hard photons, resulting in the observed soft spectrum.

Throughout the X-ray observation period (2006-2021), we found that the Compton cloud remained optically thin. However, the spectral index $(\Gamma)$ exhibited systematic variations. The hardest spectra were observed in the 2006 (XMM) observation, followed by a transition to a softer state. The softest state was observed in the 2021 (XRT4) observation. The variations in $\Gamma$ are summarized in Table~\ref{tab:spec_fit_Nthcomp} and visualized in Figure~\ref{fig:parameter}.

 As the source transitioned from one state to another state, the other spectral parameters, such as the accretion rates, the size of the Compton cloud, the temperature of the Compton cloud etc, changed accordingly. We found that the disc accretion rate $(\Dot{m}_d)$ increased from $0.051\pm0.01$ $\dot{M}_{Edd}$ (in Table~\ref{tab:spec_fit_TCAF}) to $0.08\pm0.01$ $\Dot{m}_{Edd}$ as the spectra became softer. On the other hand, the halo accretion rate $(\Dot{m}_h)$ increased from $0.07\pm0.01$ $\Dot{m}_{Edd}$ to $0.09\pm0.01$ $\Dot{m}_{Edd}$ before reaching a relatively constant value of $\sim0.1\pm0.01$ $\Dot{m}_{Edd}$.
 
The change in $\Dot{m}_d$ also influenced the position of the shock location $(X_s)$. As the disc rate increased and the source transitioned from a harder to a softer state, the shock location moved inward. This resulted in a decrease in the size of the Compton cloud, a behavior commonly observed in AGNs and Galactic black holes \citep{CT1995, Nandi2019, Nandi2021}. The shock strength $(R)$ showed relatively minor variations, ranging from $1.34\pm0.06$ to $1.92\pm0.99$, with an average value of $1.69\pm0.59$.

In addition, We also estimated the hydrogen column densities $(N_H)$ along the line of sight, while assuming a constant galactic hydrogen column density. The results showed that the extragalactic hydrogen column densities varied within a range of approximately $(10^{19}-10^{20})$/cm$^2$. This result supports that the source has a `bare-nucleus', as previously reported by \citet{Walton2013}. Furthermore, we conducted an estimation of the mass of the central supermassive black hole (SMBH) using {\tt TCAF} fitting. The obtained value for the SMBH mass was $(7.13\pm1.23)\times10^5$ M$_\odot$. This measurement aligns with the results obtained by \citep{Bentz2016} ) using an independent method, further confirming the consistency of our findings.

\subsection{Evolution of Soft Excess}

The soft excess \citep{Arnaud1985, Singh1985, Fabian2002, Gierlinski2004},  characterized by an excess emission below 2.0 keV, is an extraordinary feature in the X-ray spectra of most Seyfert 1 AGNs. Despite its common occurrence, the origin of this excess emission remains unclear. However, multiple explanations are proposed to explain this feature in literature \citep{Miniutti2004, Crummy2006, Sobolewska2007, Done2007,  Mehdipour2011, Done2012, Lohfink2012, Walton2013, Liebmann2018}. In this context, a long-term X-ray observation of UGC~6728 showed a strong correlation between the primary continuum and the soft excess \citep{Nandi2023}. In this study, we also observe the presence of the soft excess. However, the strength of this component exhibited variability throughout our analysis.

\begin{figure} 
	\centering
	\includegraphics[scale=0.7]{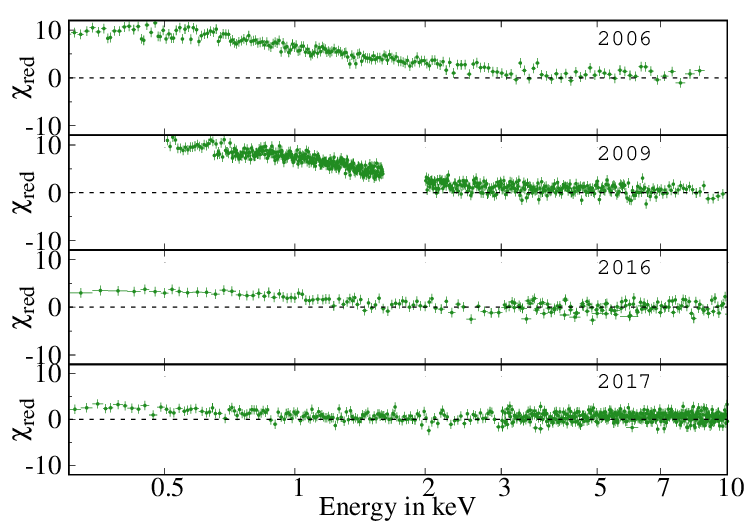}
	\caption{ The variation of the strength of soft excess with respect to the primary continuum is plotted for the observations from 2006 to 2017. The trend indicates a decline in the strength of the soft excess as the source moved from a relatively softer state to a comparatively harder state.}
	\label{fig:sx} 
\end{figure}

The presence of a strong soft excess is observed in 2006 (XMM) and 2009 (SU) observations as presented in the top two panels of Figure~\ref{fig:sx}. During these observations, the spectra appear relatively harder with  $(\Gamma=1.39^{+0.10}_{-0.11}~\text{and}~\Gamma=1.50^{+0.06}_{-0.07})$ compared to observations performed in later epochs. The electron cloud exhibited temperatures of $kT_e=470.54^{+145.62}_{-148.26}$ and $kT_e=368.87^{+103.08}_{-98.65}$ keV and optical depth $\tau=0.75^{+0.42}_{-0.35}$ and $\tau=0.61^{+0.32}_{-0.31}$, respectively. Given that the source was in a soft state during these observations, a larger Compton cloud was observed, with radii of $X_s=31.58^{+4.65}_{-3.56}$ $r_g$ and $X_s=29.09^{+2.10}_{-2.09}$ $r_g$, respectively. The disc and halo accretion rates are estimated to be  $0.05\pm$ and $0.07\pm0.01$ $\Dot{m}_{Edd}$, respectively, for these two observations. The increased size of the Compton cloud could produce a relatively cooler electron region near the outer boundary of CENBOL or Compton cloud, leading to the enhanced production of photons in the soft energy band (below 3.0 keV), as described in \cite{Nandi2021}. As a result, a strong presence of soft excess could be observed during the initial epochs.

The soft excess component exhibited a diminished presence in the 2016 (XRT1+N1) and 2017 (XRT2+N2) observations, as illustrated in the bottom panels of Figure~\ref{fig:sx}. In comparison to the preceding XMM and SU observations, these spectra appeared softer. From the spectral fitting by the TCAF model, we estimated the size of the Compton cloud for these observations were $X_s=23.62^{+2.55}_{-2.56}$ $r_g$ and $X_s=21.13^{+2.07}_{-1.56}$ $r_g$ respectively. The reduction in the size of the Compton cloud led to a decreased number of photons being generated in the primary continuum as well as a weaker soft excess component in these observations. On the other hand, we also observed weaker reflection fraction ($R_{refl}$) for weaker soft-excess \citep{Fabian2002}. It would be interesting to observe UGC 6728 in the future to see whether the soft excess diminishes or re-brightens, preserving the primary continuum \cite{Mehdipour2023}.

\subsection{Correlation between Spectral Parameters}

\begin{table}
\centering
    \caption{Correlation among different spectral parameters estimated from the spectral fitting of long term $(\sim15 \text{ years})$ X-ray observations with various satellites. The correlation analysis involved Parameter-1 and Parameter-2, which are displayed in the first and second columns, respectively. The third column represents the Pearson Correlation Coefficient (PCC), while the corresponding $p$-values are presented in the final column. A positive PCC value indicates a positive correlation between the two parameters, whereas a negative value indicates an anti-correlation.}    
    \label{tab:corr}
    \begin{tabular}{|c|c|c|c}
    \hline
Parameter-1 & Parameter-2 & PCC & $p-value$ \\
    \hline                                                
$\Gamma$ & $kT_e$ & $-0.97$ & $<0.05$ \\
$\Gamma$ & $\Dot{m}_d$ & $+0.88$ & $<0.05$ \\
$\Gamma$ & $\Dot{m}_h$ & $+0.66$ & $0.15$ \\
$\Gamma$ & $\Dot{m}$ & $+0.87$ & $<0.05$ \\
$\Gamma$ & $X_s$ & $-0.96$ & $<0.05$ \\
$\Gamma$ & $R$ & $-0.24$ & $0.64$ \\
$\Dot{m}_d$ & $X_s$ & $-0.94$ & $<0.05$ \\
$\Dot{m}_d$ & $\Dot{m}_h$ & $-0.59$ & $0.22$ \\
$\Dot{m}_d$ & $R$ & $+0.06$ & $0.91$ \\
$\Dot{m}_h$ & $X_s$ & $-0.52$ & $0.29$ \\
$\Dot{m}_h$ & $R$ & $-0.54$ & $0.28$ \\
     \hline
    \end{tabular}
\end{table}

The spectral analysis using {\tt Nthcomp} and {\tt TCAF} models reveals the nature of the region where the X-ray photons are generated. For UGC~6728, we found that the Compton cloud remains optically thin throughout the entire observation period (see Figure~\ref{tab:spec_fit_Nthcomp}). However, we observe a transition of the source from a comparatively hard state ($\Gamma=1.38$) to a comparatively soft state ($\Gamma=1.85$) between 2006 and 2021. The X-ray luminosities ($0.5-10.0$ keV) remain relatively constant, around $L_x\sim10^{42.15}$ erg/s. 

\begin{figure} 
	\centering
	\includegraphics[scale=0.7]{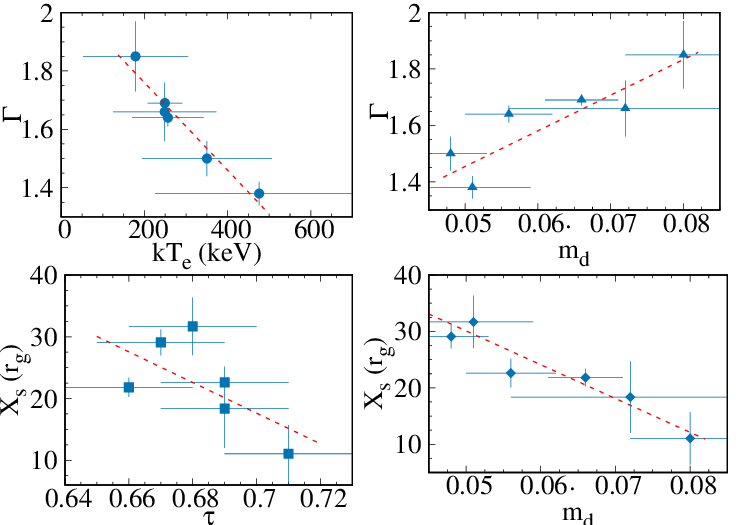}
	\caption{Correlation between different spectral parameters estimated from the spectral fitting of X-ray spectra.}
	\label{fig:par_corr} 
\end{figure}

We examined the correlation among various spectral parameters and the corresponding result is presented in Table~\ref{tab:corr}, and selected correlations are plotted in Figure~\ref{fig:par_corr}. The Pearson Correlation Coefficient (PCC) is utilized to examine the relationships between parameters\footnote{\url{https://www.socscistatistics.com/tests/pearson/default2.aspx}}. We observe a strong anti-correlation between the spectral index ($\Gamma$) and the temperature of the Compton cloud ($kT_e$) with a PCC value of $-0.97$. This anti-correlation is commonly observed in AGNs \citep{Nandi2021}, and it is displayed in the upper left panel of Figure~\ref{fig:par_corr} with linear interpolation. However, due to the inability to constrain $kT_e$ accurately with the available spectra up to $79.0$ keV, we obtained high $p$-values during the calculation of the correlation coefficient. Consequently, correlations involving $kT_e$ are not presented in Table~\ref{tab:corr}.

Next, we found a strong correlation between $\Gamma$ and the disc accretion rate ($\Dot{m}_d$) with a PCC of $0.88$ ($p$-value = $0.051$). As the disc rate increases, more disc (seed) photons are generated, leading to interactions with the Compton cloud and a subsequent decrease in cloud temperature. Consequently, a smaller number of hard photons are produced, resulting in the transition of the source from a hard to a soft state, as indicated by an increase in $\Gamma$. We also plotted the variation of $\Gamma$ and $\Dot{m}_d$ with linear interpolation in the upper left panel of Figure~\ref{fig:par_corr}. The parameter $\Gamma$ is also loosely correlated with the halo accretion rate $\Dot{m}_h$ with PCC=$0.66$. However, the high $p$-value ($p$-value = $0.15$) makes it challenging to draw any definitive conclusions based on this correlation coefficient.

Furthermore, we calculated the correlation between $\Gamma$ and the total accretion rate $\Dot{m}$, and the corresponding PCC is estimated to be  $0.88$ with $p$-value=$<0.05$. This correlation is well-established and commonly observed in AGNs and Galactic black holes. We observed a strong global anti-correlation trend between $\Gamma$ and $X_s$, with a Pearson Correlation Coefficient (PCC) value of $-0.96$ and corresponding $p$-value of $<0.05$. This finding suggests that as the shock location or the boundary of the Compton cloud ($X_s$) decreases, the spectrum tends to soften, aligning with the concept described by \citet{CT1995}. In line with this understanding, we can also anticipate an anti-correlation between $X_s$ and $\tau$, as illustrated in the lower left panel of Figure~\ref{fig:par_corr}.

In our investigation of the accretion flow parameters estimated from the spectral fitting using {\tt TCAF} model, we explored correlations among $\Dot{m}_d$, $\Dot{m}_h$, $X_s$, and $R$. It is generally anticipated that the shock location moves inward as the disc accretion rate increases \citep{CT1995, Chakrabarti1995}. We discovered a similar correlation between $\Dot{m}_d$ and $X_s$, exhibiting an anti-correlation with a Pearson Correlation Coefficient (PCC) of $-0.94$ and a corresponding $p$-value of $<0.05$. The relationship between $\Dot{m}_d$ and $X_s$ is depicted with linear interpolation in the right panel of Figure~\ref{fig:par_corr}. As for the other parameters, we observed weak correlations or anti-correlations among them. This outcome is expected since these parameters are intrinsic to {\tt TCAF} and act independently during X-ray spectral fitting.

\section{Conclusions}
\label{sec:conclusions}

We conducted a comprehensive analysis of X-ray observations spanning approximately 15 (2006-2021) years for UGC~6728, a low-mass Seyfert~1 AGN  which has a highly spinning black hole at the centre with the inclination angle nearly $\sim50$ degrees. Throughout this period, we observed significant variations in the X-ray luminosity of the source. The X-ray luminosity exhibited fluctuations on a yearly timescale, with a factor of approximately 2.4 variations observed in the 2016-2017 observations. Our investigation focused on the broadband X-ray spectra covering the energy range of 0.3-79.0 keV. We find this source is retaining its `bare'-nature throughout the observational period. In this section, we present a summary of our major findings. 


\begin{enumerate}

  \item[1.] Based on our spectral analysis of the long-term X-ray data of UGC~6728, we determined that the variations in luminosity are not dependent on the hydrogen column density ($N_H$). This outcome is consistent with the expectation that this source possesses a negligible $N_H$ value ($N_H \leq 1.0 \times 10^{19}, \text{cm}^{-2}$) along the line-of-sight.
  
  \item[2.] Throughout the extended observation period of approximately 15 years (2006-2021), the spectral slopes of UGC~6728 exhibit significant variations. Notably, this undergoes a systematic transition from a comparatively harder state $(\Gamma\sim1.38)$ to a comparatively softer state $(\Gamma\sim1.85)$ from 2006 to 2021.  

  \item[3.] The observed variation of $\Gamma$ can be attributed to the accretion flow dynamics surrounding the central supermassive black hole and the characteristics of the Compton cloud. Our analysis reveals that the disc rate $(\Dot{m}_d)$ increases from $0.05\pm0.01$ $\dot{M}_{Edd}$ to $0.08\pm0.01$ $\dot{M}_{Edd}$, while $\Gamma$ changes from $1.38\pm0.14$ to $1.85\pm.0.05$, indicating a transition of this source from a hard state to a soft state. Correspondingly, the shock location $(X_s)$ or the size of the Compton cloud and compression ratio $(R)$ exhibit variations. Notably, the Compton cloud is optically thin $(\tau<1.0)$ throughout the entire observation period. 

  \item[4.] The intensity of the soft excess component varied over the duration of our observational period. The 2006 and 2009 observations displayed profound soft excess, compared with the diminished presence of soft excess observed in 2016 and 2017 epochs. The variation in the strength of soft excess could be explained by the variation of the accretion dynamics.

  \item[5.] In our timing study, we observed a similar order of variability in the high energy band (above $3.0$ keV). Furthermore, correlation analysis between two high energy bands, specifically $3.0-6.0$ keV and $7.0-79.0$ keV, does not reveal any time delays, suggesting that the photons in both energy ranges may have originated from a similar mechanism. Since the high energy spectra (above $3.0$ keV) of UGC~6728 can be adequately described by a single power-law model, we can infer that this source exhibits either no reflection hump or a very minimal hump in relation to the primary continuum above $10.0$ to $15.0$ keV.

  \item[6.] By performing X-ray spectral fitting using {\tt TCAF} model, we determined the mass of the central supermassive black hole (SMBH) to be $M_{BH}=(7.13\pm1.23)\times10^5$ M$_\odot$ with a high spin at $a=0.97^{+0.20}_{-0.27}$ and inclination angle of $49.5\pm14.5$ degree.

\end{enumerate}

\section*{Acknowledgements}
\label{sec:ack}
 We would like to express our sincere gratitude to the reviewers for their valuable insights and comprehensive evaluation of our work, greatly contributing to its improved quality and clarity. This research work at PRL is funded by the Department of Space, Government of India. AC and SSH are supported by the Canadian Space Agency (CSA) and the Natural Sciences and Engineering Research Council of Canada (NSERC) through the Discovery Grants and the Canada Research Chairs programs. This research has made use of data and/or software provided by the High Energy Astrophysics Science Archive Research Center (HEASARC), which is a service of the Astrophysics Science Division at NASA/GSFC and the High Energy Astrophysics Division of the Smithsonian Astrophysical Observatory. This work has made use of data obtained from the {\it NuSTAR} mission, a project led by Caltech, funded by NASA and managed by NASA/JPL, and has utilized the {\tt NuSTARDAS} software package, jointly developed by the ASDC, Italy and Caltech, USA. This research has made use of observations obtained with {\it XMM-Newton}, an ESA science mission with instruments and contributions directly funded by ESA Member States and NASA. This work made use of {\it Swift}/XRT data supplied by the UK Swift Science Data Centre at the University of Leicester, UK.

\section*{Data Availability}

We used archival data of {\it Swift}/XRT and {\it NuSTAR} observatories for this work. These data are publicly available on their corresponding websites. Appropriate links are given in the text.






\bibliographystyle{mnras}
\bibliography{references} 







\bsp	
\label{lastpage}
\end{document}